\documentclass[aps,preprint,superscriptaddress,showpacs]{revtex4}
\usepackage[english]{babel}
\usepackage{epsfig}
\usepackage{graphics}
\usepackage{natbib}
\usepackage{amsthm}
\usepackage{amsmath}
\usepackage{times}
\usepackage{psfrag}
\usepackage{subfigure}
\usepackage{epstopdf}

\usepackage{soul}

\begin{document}

\title{Can L\`evy noise induce coherence and stochastic resonances in  a birhythmic van der Pol system?}

\author{\textbf{R. Mbakob Yonkeu}}
\affiliation{Fundamental Physics Laboratory, University of Bamenda,
Faculty of Sciences, Department of Physics, PO Box 39, Bambili, NWR, Cameroon.}
\author{\textbf{R. Yamapi}}
\email[Author to whom correspondence should be addressed. Electronic mail:]{ryamapi@yahoo.fr}
\affiliation{Fundamental Physics Laboratory, Physics of Complex System group,
Department of Physics, Faculty of
 Science, University of Douala, Box 24 157 Douala, Cameroon.}
\author{\textbf{G. Filatrella}}
 \affiliation{INFN Gruppo collegato Salerno and Department  of Sciences and Technologies
\small University of Sannio, Via F. De Sanctis,
I-82100 Benevento, Italy.}
\author{\textbf{J\"urgen Kurths}}
\affiliation{Potsdam Institute for Climate Impact Research (PIK), 14473 Potsdam, Germany}
\affiliation{Department of Physics, Humboldt University, 12489 Berlin, Germany}
\date{\today}

\begin{abstract}
The analysis of a birhythmic modified van der Pol type oscillator driven by periodic excitation and L\`evy noise shows  the possible occurrence of coherence resonance and stochastic resonance.
The frequency of the harmonic excitation  in the neighborhood of one of the limit cycles
 influences the coherence of the dynamics on the time scale of intrawell oscillations.
The autocorrelation function, the power spectral density and the signal-to-noise-ratio used in this analysis are shown to be maximized for an appropriate choice of the noise intensity.
A  proper  adjustment   of the L\`evy noise intensity enhances the output power spectrum of
  the system, that is, promotes  stochastic resonance.
Thus, the robustness of the resonance, that seems to occur also in the presence of nonstandard noise,  is examined using  standard measures.
The initial selection of the attractor seems to have an influence on the coherence, while the skewness parameter of the L\`evy noise has not a notable   impact on the resonant effect.

\textbf{Keywords: Stochastic resonance, L\`evy noise, birhythmic system.}

\end{abstract}
\pacs{\\
 02.50.Ey - Stochastic processes \\
05.40.-a Fluctuation phenomena, random processes, noise, and Brownian motion \\
05.10.Gg - Stochastic analysis methods (Fokker-Planck, Langevin, etc.).}


\maketitle

\newpage

\textbf{}

\section{Introduction}
\label{introduction}

Stochastic dynamical systems arise as mathematical models for complex phenomena in almost all scientific areas, for the presence of noise, due to internal or external fluctuations, is inevitable in real world systems.
Special phenomena such as stochastic resonance (SR) \cite{pikovsky,Kenfack,Alfonsi},  noise-induced transitions \cite{Gu,Zhao}, noise enhanced stability \cite{Fiasconaro05,Guarcello15}, and stochastic bifurcations \cite{Zakhrova,Mbakob2,Mbakob3,Chamgoue,Biswas19} have attracted a great deal of attention in various fields including physics, chemistry, biology, and engineering.
SR, a term coined by Benzi et al. \cite{Benzi} to explain the cyclical variation of the warm  and the cold climate in paleo-climatology, has been broadly applied to describe a counter-intuitive phenomenon where the presence of a suitable amount of noise can optimize the output signal quality in a nonlinear system \cite{Mcdonnell,Gammaitoni}.

Depending on the applications in science and engineering, noise can be grouped into two main  broad categories: Gaussian noise (GN) and non-Gaussian noise.
GN is just an ideal model that cannot describe random fluctuations with finite variance, while non-GN, widely observed in various areas such as biology, seismology,
electrical engineer, and finance \cite{H.Qiao}, is characterized by large, potentially infinite, jumps.
 L\`evy noise (LN) is a class of stable non Gaussian noise that exhibits long heavy tails of its distribution.
The archetypal phenomena of noise-induced ordering are robust and can be detected
also in systems driven by Markovian, non-Gaussian, heavy-tailed
fluctuations with infinite variance \cite{Bartlomiej}, and has been investigated in diverse systems as Josephson junction \cite{Guarcello13,Valenti14,Guarcello16} and in biophysics to  help detect faint signals \cite{art,Guarcello17,Guarcello19}.
However, a LN model more accurately describes how the neuron's  membrane potential evolves than a simpler diffusion model  because the more general L\`evy model includes not only pure-diffusion and pure-jump models but jump-diffusion models as well \cite{art1}.
In fact, stochastic L\`evy processes describe transport  processes with anomalous diffusion, as characterized by an anomalous   mean squared displacement ($\Delta x$ represents the independent increment),\emph{ i.e.},
 \begin{equation}
\label{eq1a}
\left< \left( x(t) - \langle x(t) \rangle \right)^{2} \right> = \langle (\Delta x)^{2} \rangle \propto t^{\kappa},
\end{equation}
\noindent with $\kappa = 3-\gamma \neq 1.0$, at variance with the linear dependence $\langle (\Delta x)^2 \rangle \propto t$ that characterizes Gaussian  diffusion, is connected with the LN index $\gamma$.
Therefore, $\kappa$ determines the anomalous diffusion exponent that specifies the process,  either sub-diffusive (with $0<\kappa<1$), super-diffusive (with $\kappa>1.0$), whereas the ordinary Gaussian case corresponds to the ballistic case (with $\kappa=2.0$) \cite{B.D,P.M}.
Hence, LN can be considered as a standard stochastic process that describes in the simplest fashion the effect of fast surroundings \cite{Anishchenko}.
However, compared with GN, LN is frequently encountered in nature, since long jumps are associated with a complex structure of the environment \cite{T.Srokowski}.
L\`evy stable distributions are a rich class of probability distributions and have many intriguing mathematical properties.

LN driven birhythmic systems are known to manifest interesting physical properties.
The response of birhythmic systems to noise is peculiar, in that the system possesses two stable attractors, each characterized by  two frequencies.
Recently, it has been shown  that LN can induce transitions between attractors and enhanced stability in a birhythmic van der Pol system \cite{yamapi2019}.
The theory of escape time from one attractor to another quantifies the overall stability of the attractors of this system, also in the presence of LN.
Numerical simulations have demonstrated that in the presence of LN, the induced escapes from an attractor to another are similar to the escapes between stable points in an ordinary potential.
Comparing the escapes under the influence of LN with the Gaussian case, it is evident that the differences are more pronounced for large values of the L\`evy index $\gamma$.
In this work we consider both the effects of periodic signal and L\`evy noise in the birhythmic self-sustained system.
The present aim is to check if the L\`evy noise can induce CR and SR.
The occurrence of SR phenomenon implies that these measures will
exhibit a well-marked maximum at a particular noise level \cite{Dybiec}.
The CR and SR can be determined by using various measures, including
residence time distributions, Power Spectrum Density (PSD),
Signal-to-Noise  Ratio (SNR), spectral power amplification, input/output
cross-correlation measures, and probability of detection \cite{Reinker}.
In this paper we will use some measures to hilight different effects of LN on birhythmic systems.

The paper is organized as follows. Section \ref{driven} describes the birhythmic van der Pol system driven by LN and periodic signal, and the algorithm of the numerical simulations.
This section start by recalling the birhythmic properties on the free birhythmic van der Pol system, and the parameter region where birhythmicity appears; the L\`evy process and the algorithm of numerical simulations are also presented.
Section \ref{tools} deals with the diagnostic tools to detect coherence and stochastic resonances between noise and deterministic oscillations in the stochastic birhythmic van der Pol system. The purpose is to formalize the main tools employed in this work.
Section \ref{coherence} studies coherence and stochastic resonances-like phenomena in the birhythmic van der Pol system.
By means of the numerical tools it is possible to quantify the degree of coherence (or anti-coherence), as described in Sect. \ref{coherence}.
In the same section is also discussed the occurrence of  SR for some sets of parameters, which depend highly on the noise intensity.
The last Section presents the conclusions.

\section{The driven birhythmic van der Pol system with L\`evy noise and numerical description}
\label{driven}

This Section is devoted to the model description and to summarize the
 main physical properties of the birhythmic van der Pol type oscillator.

\subsection{The driven birhythmic van der Pol system with L\`evy noise}

\begin{figure}
\begin{center}
\includegraphics[height=6.4cm,width=12.4cm]{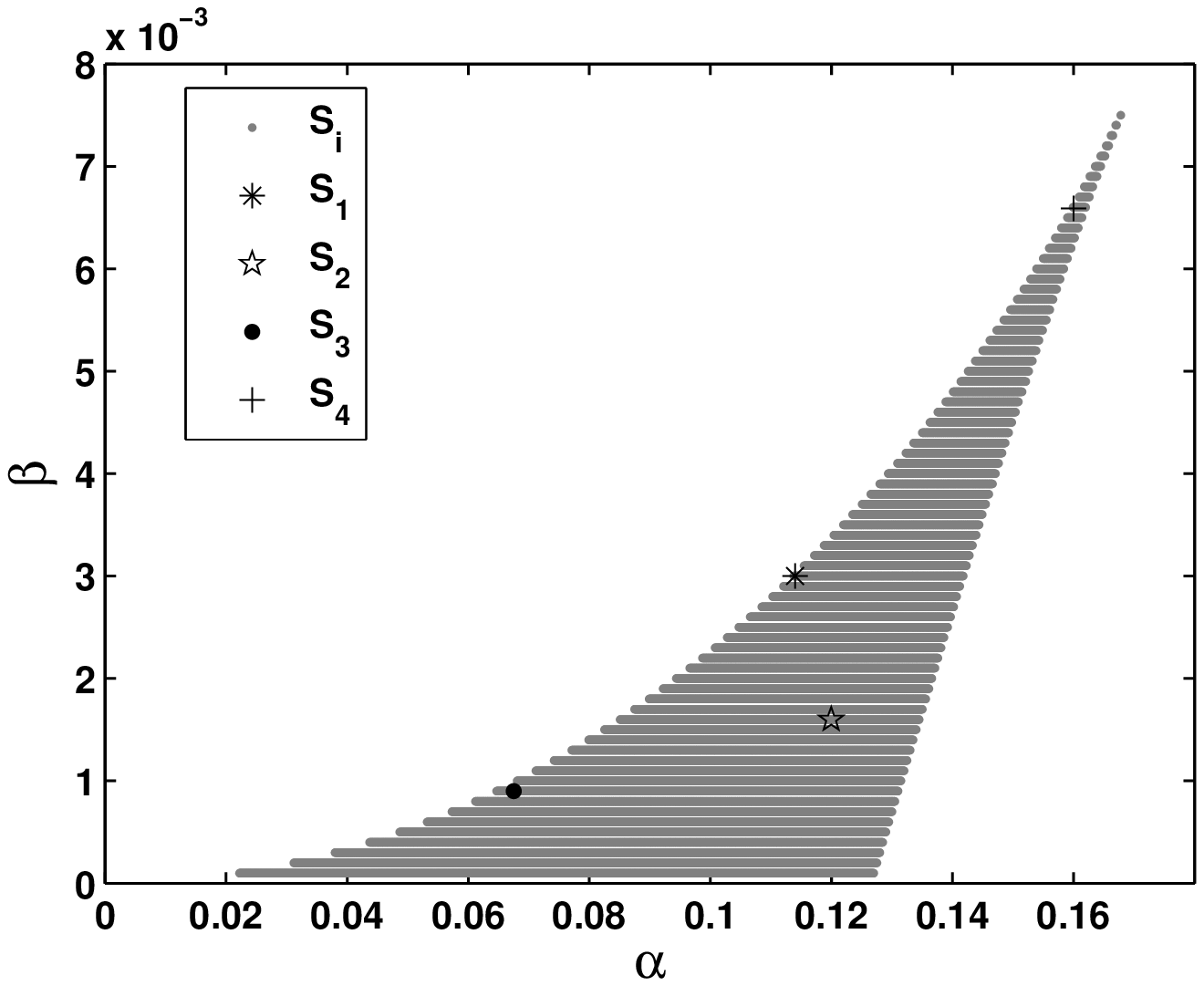}
\caption{\it
Parameter region of  single limit cycle (white area) and three limit cycles (gray area) with $\mu=0.01,$
 obtained from simulations of Eq.(\ref{eq2}). $S_i$ refers to Table \ref{Tab1}.}
\label{fig1}
\end{center}
\end{figure}

The system under study is a model which can be used to describe, for instance, a biological  enzyme-substrate system with a ferroelectricity behavior in brain wave \cite{Chamgoue}.
Assuming that the influence on the enzymes amounts is a combination of random (LN type) and periodic excitation, it appears  that the stochastic evolution equation of the activated enzyme molecules, on the physical description used by Fr\"ohlich \cite{Frohlich1,Frohlich2} can be mapped onto the following stochastic driven well-known variation of the van der Pol equation \cite{Yamapi10,Ghosh11,Guo19}:
\begin{equation}
\label{eq1}
\ddot{x}-\mu(1-x^{2}+\alpha x^{4}-\beta x^{6})\dot{x}=\Gamma(t)+\eta(t),
\end{equation}
in the enzymatic-substrate reactions interpretation of Eq.(\ref{eq1}) $x$ is proportional to the population of enzyme molecules in the excited polar state, $\Gamma(t)=E\sin\omega t$ is the periodic drive, $E$ and $\omega$ are the amplitude and frequency, respectively.
The random term $\eta(t)$ denotes the LN generalized \emph{Wiener} process, obeying the L\`evy distribution $L_{\gamma, b}(\eta,\sigma,\mu_1)$.
In fact, $\eta(t)$ is the time derivative of the L\`evy process $\zeta$.
The overdot denotes the derivative with respect to time, $\alpha$ and $\beta$ are positive parameters which indicate the system behavior to a ferroelectric instability compared with its electrical resistance and $\mu$ is a non linear parameter that effectively refers to the strength of nonlinear damping \cite{Enjieu}.
Equation (\ref{eq1}) without LN and periodic force is a multi-limit cycle oscillator \cite{Yamapi10}:
\begin{equation}
\label{eq1a}
\ddot{x}-\mu(1-x^{2}+\alpha x^{4}-\beta x^{6})\dot{x}=0.
\end{equation}
The oscillator described by Eq. (\ref{eq1a}) exhibits self-sustained oscillations possessing more than one limit cycle, which is the condition for  birhythmicity.
Following Ref.\cite{Enjieu}, the periodic solutions of Eq.(\ref{eq1a}) can be approximated by
\begin{equation}
\label{eq2}
x(t)=Asin(\Omega t),
\end{equation}
where $A$ represent the amplitude and $\Omega$ is the frequency
of the system without periodic force and LN.
The amplitude equation is as follows \cite{Enjieu}:
\begin{equation}
\label{eq3}
\frac{5\beta}{64}A^{6}-\frac{\alpha}{8}A^{4}+\frac{1}{4}A^{2}-1=0,
\end{equation}
which determines a codimension two saddle-node bifurcations.
The frequency of the oscillations in the minimum $\Omega$ is approximately given by:
\begin{eqnarray}
\label{eq21}
\Omega=1+\mu^2\Omega_2+o\left(\mu^3 \right)
\end{eqnarray}
where
\begin{center}
$\Omega_2=\frac{93\beta^2}{65536}A^{12}-\frac{69\alpha\beta}{16384}A^{10}+
\left( \frac{67\beta}{8192}+\frac{3\alpha^2}{1024} \right) A^8
-\left(\frac{73\beta}{2048}+\frac{\alpha}{96}\right) A^6
+\left( \frac{1}{128}+\frac{\alpha}{24}\right)A^4
- \frac{3}{64}A^2.$
\end{center}
Figure \ref{fig1} describes the bifurcation lines that enclose the region  of birhythmicity in the two parameters domain $(\alpha, \beta)$.
In the gray region are observed three limit cycles (and hence the system is birhythmic for the presence of two stable orbits), while one limit cycle appears in the white region \cite{Mbakob1,Mbakob2}.

The orbits' stability can be characterized by means of a pseudo-potential function $U(A)$\cite{Mbakob3}.
In the absence of noise and in the presence of a harmonic drive, the pseudo-potential can approximated as follows:

\begin{equation}
\label{eq3b}
 -U(A)=\frac{\mu}{128}
  \left[ 32 A^{2}-4A^{4}+\frac{4}{3}\alpha A^{6}-\frac{5}{8}\beta A^{8} \right]
 - \frac{\mu E}{2\omega}A .
\end{equation}

\begin{figure}
\begin{center}
\includegraphics[height=6.4cm,width=7.4cm]{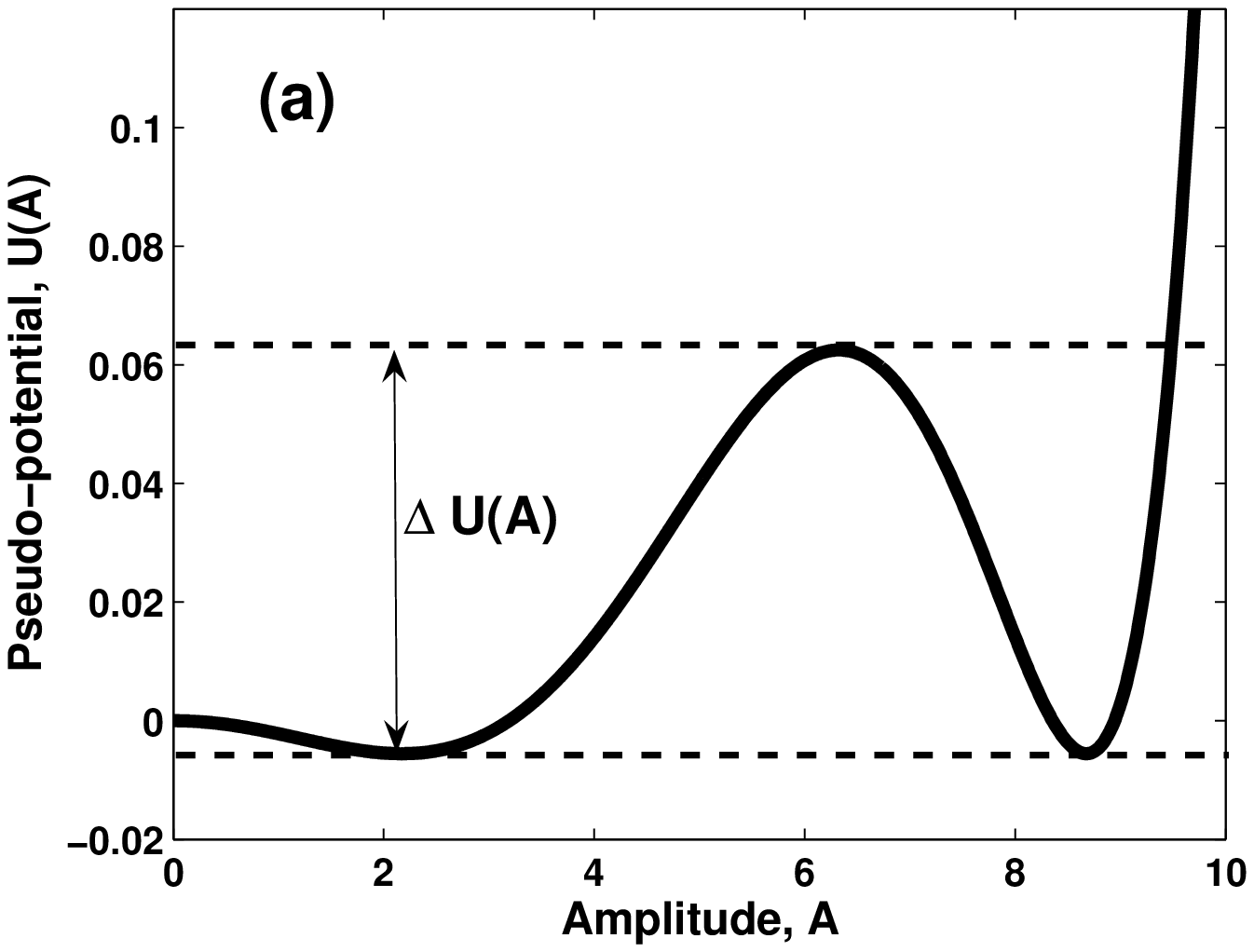}
\includegraphics[height=6.4cm,width=7.4cm]{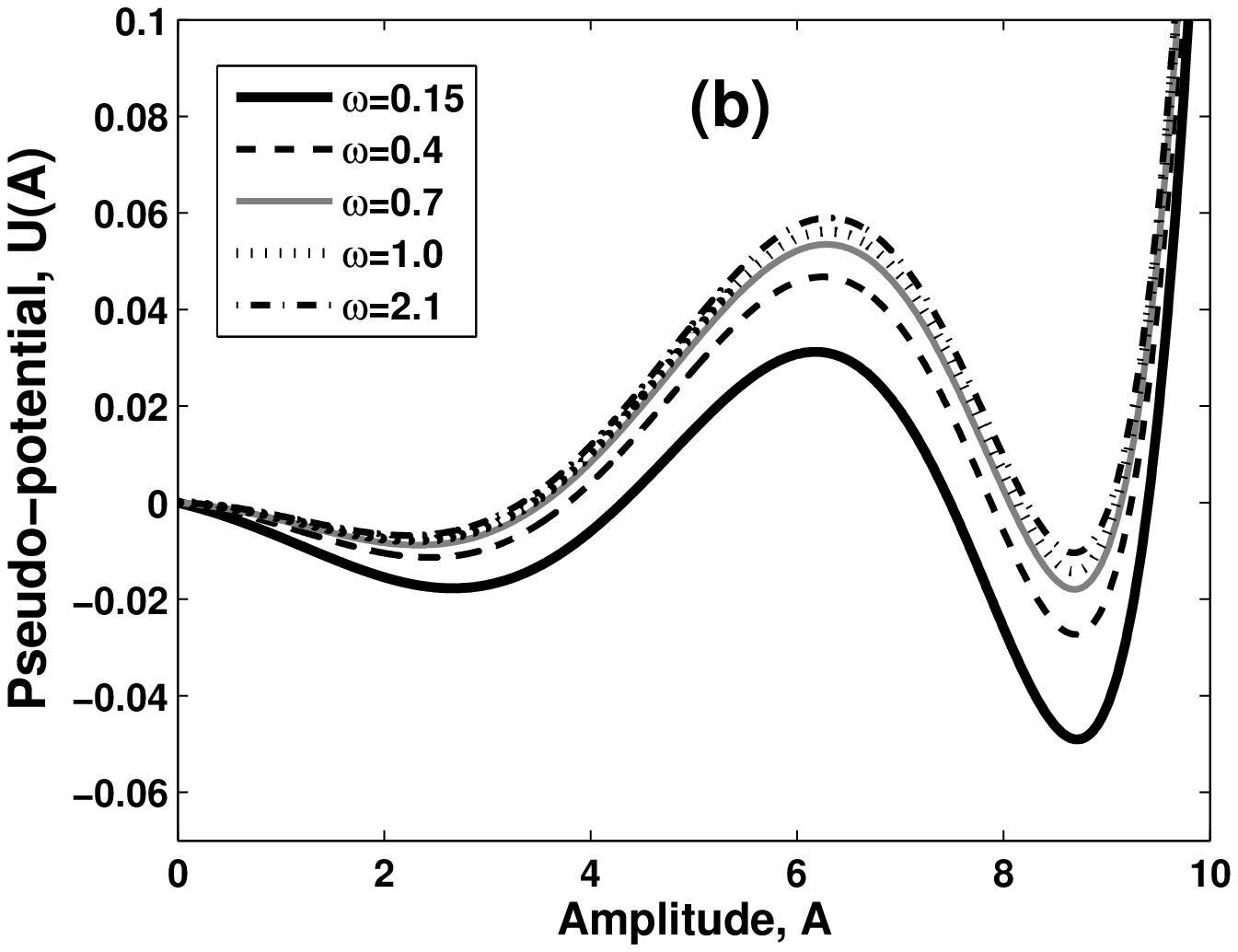}
\caption{\it
The generic of pseudo-potential $U$ versus the amplitude $A$
with $\alpha=0.0675$, $\beta=0.0009$, $\mu=0.01$,
$t={\pi}/{2}$, $E=0.0$ left panel and $E=0.2$ right panel.}
\label{potentiel}
\end{center}
\end{figure}

The pseudo-potential (or quasi-potential) is shown in Fig.\ref{potentiel} for parameters $\alpha$ and $\beta$ such that it appears almost symmetrical for $E=0.0$ (Fig.\ref{potentiel}(a)) or $E=0.2$ (Fig.\ref{potentiel}(b)) as a function of the amplitude for
several values of the frequency of the applied drive.
It appears that the effect of periodic force is still visible also for $\omega=1.5$
for $E =0.2$ and when $\omega>1.0$, its effect is no longer visible. The periodic
force destroys the symmetry of the potential.
In the interpretation of the pseudo-potential as a {\it bona fide} potential, the
system appears to be alike to an ordinary bistable system, which offers the possibility for a particle to periodically roll from one potential well to the other.
In the birhythmic  region four cases have been selected, as reported in Table \ref{Tab1},
 in which \cite{Mbakob1}: i) the potential is asymmetric and the frequencies
 are almost identical ($S_1$) , ii) the potential is asymmetric but  the
 frequencies are different ($S_2$), iii) and iv) the potential is
 symmetric and the frequencies are almost identical ($S_3$ and $S_4$).

\begin{table}
  \begin{center}
  \begin{scriptsize}
\begin{tabular}{|l|l|l|l|}
\hline
$S_i=(\alpha,\beta)$& Amplitudes $A_i$ & Period $P_i$ of the orbit  &Energy barriers $\Delta U_{1,3}$ \\
\hline
                   &$A_1$=2.37720                    & $P_1=6.271$     &$\Delta U_1=1.063\times10^{-2}$\\

$S_1=(0.114;0.003)$& $A_2$=5.02638                   & unstable           &\\

                 & $A_3$=5.46665                     & $P_3=6.271$      &$\Delta U_3=2.111\times10^{-4}$ \\
\hline
                        & $A_1$=2.481848           & $P_1$=6.2809      &$\Delta U_1=1.432\times10^{-3}$ \\

$S_2=(0.12;0.0016)$     & $A_2$=3.586374                    & unstable             & \\

                        & $A_3$=10.048778                 & $P_3=9.2352$     &$\Delta U_3=0.4424$ \\
\hline
                        & $A_1$=2.1730001                   & $P_1$=6.283       &$\Delta U_1=6.509\times10^{-2}$ \\

$S_3=(0.0675;0.0009)$     & $A_2$=6.324501                  & unstable            &  \\

                        & $A_3$=8.6760004             & $P_3$=6.283       &$\Delta U_3=6.509\times10^{-2}$ \\
\hline
                        & $A_1$=2.481848               & $P_1$=6.285   &$\Delta U_1=1.036\times10^{-5}$ \\

$S_4=(0.16;0.00658)$     & $A_2$=3.586274                    & unstable        & \\

                        & $A_3$=10.048778               & $P_3$=6.285       & $\Delta U_3=1.037\times10^{-5}$\\
\hline

\end{tabular}
  \end{scriptsize} 
  \caption{\it{Amplitudes and periods of the limit cycles, and energy barriers
  for the pseudo-potential, see Fig. \ref{potentiel}.
  Data refer to the case $\mu=0.01$.}}
  \label{Tab1}
  \end{center}
\end{table}

\subsection{The L\`evy noise process and the algorithm for numerical simulations }
\label{numerical}

L\`evy distributions are a class of probability distributions with many intriguing mathematical properties \cite{Liang Y}.
The LN is governed by four parameters \cite{Applebaum,Dubkov}: Stability
 index $\gamma$ (with $0<\gamma\leq 2.0$), skewness parameter $b$ (with $-1\leq b\leq 1$),
  mean parameter $\mu_1$ (with $\mu_1 \epsilon \Re$) and scale parameter $\sigma$ (
  with $\sigma \epsilon [0;\infty[$), thus it is more powerful in
  describing general realistic fluctuations \cite{Zheng}.
 Assume that $\zeta$ obeys to L\`evy distribution $L_{\gamma, b}(\zeta,\sigma,\mu_1)$
 whose  characteristic  function is \cite{Dybiec}:
\begin{equation}
\label{eq4}
\Phi(k)=\int_{-\infty}^{+\infty}d\eta e^{ik\eta}L_{\gamma, b}(\zeta,\sigma,\mu_1).
\end{equation}
Therefore
\begin{subequations}\label{eq5}
\begin{align}
&\Phi(k)=exp\left[i\mu_1k-\sigma^{\gamma}\mid k\mid^{\gamma}(1-ib.sgn(k)tan\frac{2}{\pi}ln|k|\right]\,\,\,\,\,\,\,\, if \,\,\,\gamma\in [0;1]\cup [1;2],\\
&\Phi(k)=exp\left[i\mu_1k-\sigma\mid k\mid(1+ib.sgn(k)\frac{2}{\pi}ln|k|\right]\,\,\,\,\,\,\,\, if\,\,\, \gamma=1.
\end{align}
\end{subequations}
Note that $ \gamma$ describes an asymptotic power law of  the  L\`evy distribution
and $sgn(k)$  represent the sign function of $k$  and is defined as follows:
$$sgn(k)=
\left\{
  \begin{array}{cll}
    +1&if&k > 0, \\
    0&if&k=0,\\
    -1&if&k < 0.
  \end{array}
\right.
$$
 $\sigma$ ($\sigma\epsilon \Re$) is the center or location
parameter which denotes the mean value of the distribution, and the mean of the
distribution exists. $D=\sigma^{\gamma}$ is the scale parameter and
noise intensity respectively \cite{Janicki}. $D$ is
the intensity of the LN and has the physical meaning
of the measure of the intensity of the random electric field.
Depending on the choice of $\gamma$, it covers GN
 (where $\gamma=2.0$) and NGN (where $0<\gamma<2.0$)
 which can model  thermal fluctuations and  non-thermal ones respectively.
 The prominent characteristic feature of the distributions
  $L_{\gamma, b}(\zeta,\sigma,\mu_1)$ is its existence of
   moments up to the order $\gamma$, \emph{i.e.} the integral:
    $$\int_{-\infty}^{+\infty}L_{\gamma, b}(\zeta,\sigma,\mu_1)\zeta^{\gamma}d\zeta$$ is finite. This
  statement results in the conclusion that the only  stable distribution possessing a second moment is the Gaussian; for all the other values of $\gamma$, the variance of a stable distribution diverges.

In  this  paper,  we  use  the  \emph{Janicki-Weron}  algorithm \cite{Dybiec}
to generate the L\`evy distribution,
with strongly depend of $\gamma$, for instance:

For $\gamma\neq1$,  the L\`evy distribution is simulated as:
\begin{equation}
\label{eq6}
\zeta=D_{\gamma,b,\sigma}=\frac{sin\left[\gamma(r+C_{\gamma,b}\right]}{\left[cos(r)\right]^{\frac{1}{\gamma}}}\left[\frac{cos(r-\gamma(r+C_{\gamma,b}))}{w}\right]^{\frac{1-\gamma}{\gamma}}+\mu_1
.
\end{equation}
With $C_{\gamma,b}=\frac{\arctan\left[b\tan\left(\frac{\pi\gamma}{2}\right)\right]}{\gamma}$
$D_{\gamma,b,\sigma}=\sigma\left[\cos\left(\arctan\left[b\tan\left(\frac{\pi\gamma}{2}\right)\right]\right)\right]^{\frac{-1}{\gamma}}.$

\begin{figure}
\begin{center}
\includegraphics[height=6.4cm,width=7.4cm]{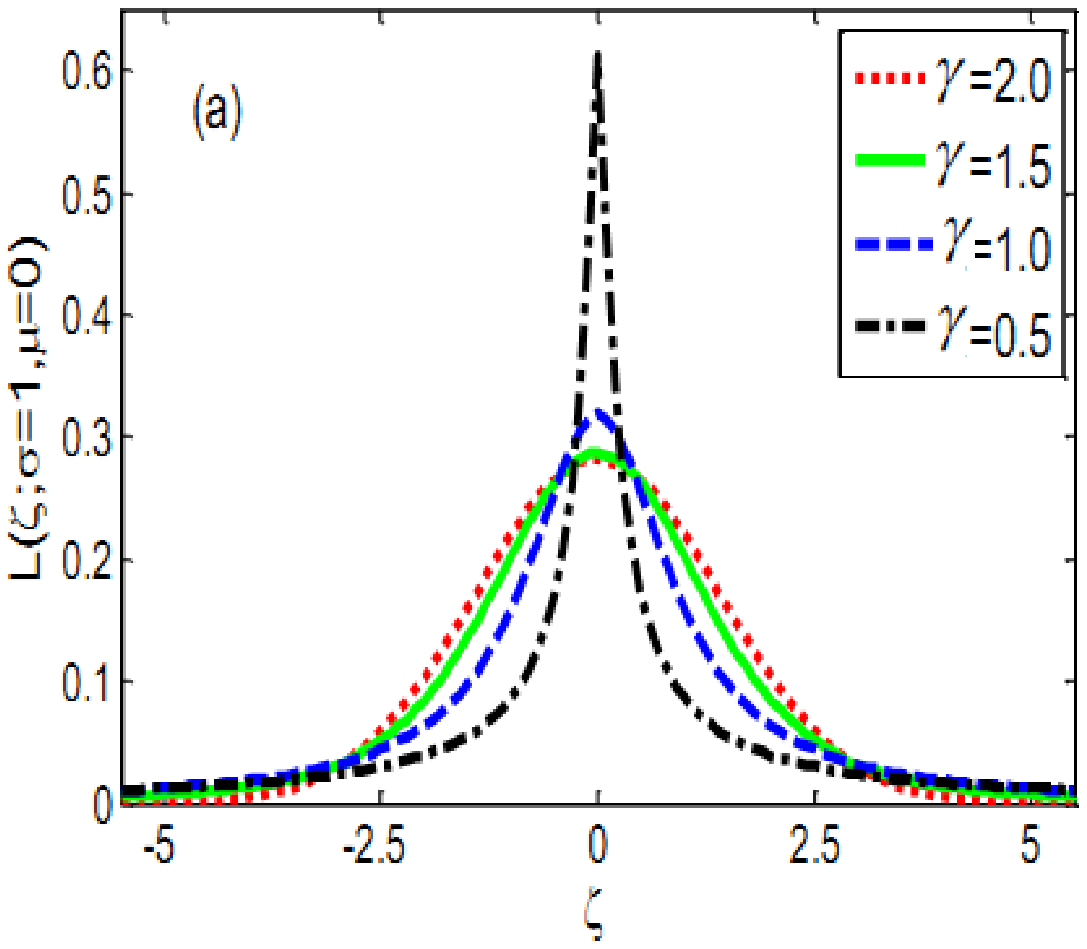}
\includegraphics[height=6.4cm,width=7.4cm]{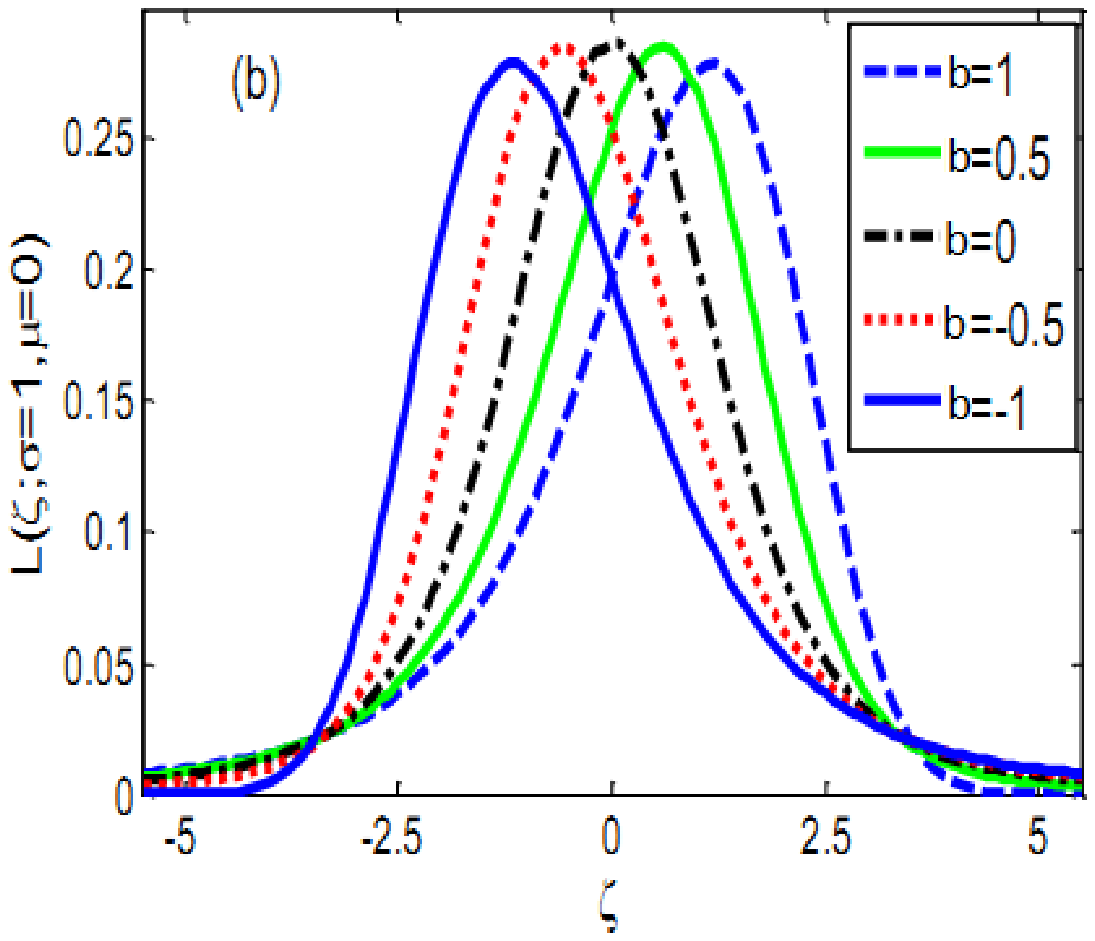}
\caption{\it
The generic of $\alpha$-stable probability function with $\mu_1=0.0$ (a) and $\gamma=1.7$ (b).
For $b=0$, distribution are symmetric, while $b\neq 0$, they are asymetric functions.}
\label{proba}
\end{center}
\end{figure}

For $\gamma=1$, it is simulated as:
\begin{equation}
\label{eq7}
\zeta=\frac{2\sigma}{\pi}\left[\left(\frac{\pi}{2}+b r\right)\tan(r)-b\ln\left(\frac{\frac{\pi}{2}w\cos(r)}{\frac{\pi}{2}+br}\right)\right]+\mu_1.
\end{equation}
Here $r$ and $w$ are independent random  variables, with $r$ uniformly distributed on $\left[-\frac{\pi}{2},\frac{\pi}{2}\right]$ and $w$ is
 standard exponential distribution. Some parameters are chosen by fixing $\mu_1=0.0$.
 In Fig.\ref{proba}, the L\`evy density function $L_{\gamma,b}(\zeta, \sigma, \mu_1)$ under $\gamma$ and $b$
       are simulated by using Janicki-Weron  algorithm \cite{Dybiec} with the parameter given as $\mu_1=0.0$, $\sigma=1.0$. From these pictures, it can be seen that the L\`evy distribution becomes symmetric when $b=0.0$ and it is reduced to the well-known Gaussian distribution when $\gamma=2$. Considering the case where $\gamma <1.0$, $L_{\gamma,b}(\zeta, \sigma, \mu_1)$ shows left-skewed with $b<0$ and right-skewed with $b>0$; The inverse phenomena are seen when $\gamma>1$. Let us underline that in practice, the noise intensity is very small, so the periodic signal will play a leading role and the noise has a
modulating action on it. All our simulations are performed with
a time step $\Delta t=0.001$ and the initial values $x(0)$ and
$\dot{x}(0)$ are taken around the stable limit cycle.

\section{TOOLS TO QUANTIFY COHERENCE AND STOCHASTIC RESONANCES}
\label{tools}

In this Section we describe some measures of the coherence of the
oscillator, to quantify the effect of noise and drive on the birhythmic oscillator.
Coherence resonance occurs in an excitable regime if noise is added and the system receives an energy that is large enough to cause excursions,
 or evasions, from the original orbit.
If, in spite of these large fluctuations the system regularity is enhanced,
 one names the phenomenon as a CR.
In the following subsections some measures of the degree of coherence are described.

\subsection{Auto-correlation function analysis}
\label{ACF}

The Auto Correlation Function (ACF) is employed to
characterize the long time-scale excursions in which transitions  can occur from an attractor to the other.
We are particularly interested to the analysis of effects of the LN index $\gamma$ on the ACF.
Noise effects are characterized by the decay rate of ACF or Cor \cite{Stratonovich}, defined as follows:
\begin{equation}
\label{eq11}
Cor(\varrho)=\frac{\left< \tilde{A}(t)\tilde{A}(t-\varrho)\right>}{\left<\tilde{A}^2\right>},
\end{equation}
where $\tilde{A}=A-\langle A \rangle $, $\varrho$ is a normalized time.
For $\varrho=0$ the ACF reads  $Cor(0)=1$.
The nonlinear behavior decreases $Cor(\varrho)$ for large $\varrho$ times, as an effect of the LN.
\begin{equation}
\label{eq12a}
\left<\tilde{A}^2\right>=\frac{1}{T^{M}}\int_0^{T^{M}}\tilde{A}^2dt
\end{equation}
The time $T^{M} \gg \varrho$ is taken equal to $10^{20}$.

To summarize the induced coherence, it is convenient to define the autocorrelation time $\tau$:
\begin{equation}
\label{eq12}
\tau=\frac{1}{100T^M}\int_0^{100T^M}Cor^{2}(\varrho)d\varrho
\end{equation}

Informally, this time characterizes the similarity between any two points in the dynamics as a function of time lag between them.
In other words,  time $\tau$ summarizes the correlation between points separated by various time lag.
Autocorrelation is a mathematical representation of the degree of similarity between a given time series and lagged version of the itself over successive time intervals.
Therefore, if there exists an optimum value of noise which enhances the coherence of the
 system response, it can be detected through the autocorrelation time.

\subsection{Power spectral density analysis}
\label{power}

The Power spectral density  is useful in the case of a short time scale, that is, in the weak noise regime, when the random disturbance is weak enough to forbid  jumps from a meta-stable state to the other in the observed time.
The power spectrum is retrieved through the numerical Fourier transform:
\begin{equation}
\label{eq13}
H(\omega)=\int x(t)e^{i\omega t}dt.
\end{equation}
The power $P(\omega)$ can in turn be computed as
\begin{equation}
\label{eq14}
P(\omega)=|H(\omega)|^{2}+|H(-\omega)|^2
\end{equation}
The discrete noise power spectrum is numerically estimated using a Fast Fourier Transform (FFT) algorithm with $n=2^{12}$ sampled data; the average power spectrum is obtained with $25000$ realizations to achieve a better accuracy.

SNR can be conveniently expressed in decibels:
\begin{equation}
\label{eq15}
SNR=10log_{10}\left(\frac{S}{N}\right),
\end{equation}
where $S$ and $N$ correspond to the average power spectra (\ref{eq14}) with an applied deterministic drive ("Signal", if  $x(t)$ is retrieved from Eq.(\ref{eq1}) with $E\neq0$), and the purely noisy system  ("Noise", if  $x(t)$ is retrieved from Eq.(\ref{eq1}) with $E=0$), respectively.
The signal $S$ and the noise $N$ values are choose around a peak in the spectrum at a frequency $\omega$.
The average power spectrum of the signal $S$ is defined as follow:
\begin{equation}
\label{eq15b}
S = \frac{1} {2\Delta\Omega}\int_{\Omega-\Delta\Omega}^{\Omega+\Delta\Omega}P(\omega)d\omega,
\end{equation}
where $\Delta\Omega$ denote a convenient interval of frequencies around the central frequency $\Omega$.
The numerical definition (\ref{eq15}) has to be adapted to the discrete nature of the numerical FFT to include in the noise $N$  the bins that contain only noise.
Moreover, to obtain reliable results, the system must explore the phase space; in particular  if the system is perturbed with very weak noise, one should ensure that the results do not depend upon the observation time or the initial conditions.


An estimate of the SNR from the power spectral density measures the response of the system to the combination of deterministic drive and noise.
A self oscillatory system is also influenced by the attractor proper
frequency.
The drive induces a sharp peak located at the drive frequency,
 emerging from the noise background and the frequency of the peak changes with the LN amplitude $\gamma$.
For a given signal, the frequency corresponding the maximum of $\mid H(\omega)\mid^{2}$ is little affected by the variation of the noise intensity $D$.
In the numerical solution of Eq.(\ref{eq1}), because of the heavy tails, discontinuity and irregular jumps of the LN may cause the numerically generated sample paths to diverge  decreasing  $\gamma$.
To circumvent this problem, some methods have been used, for example, {Zhang}
and {Song} \cite{Zhang and Song} impose a constraint on the value of the solution $x(t)$.

\section{The coherence resonance}
\label{coherence}

\begin{figure}
\begin{center}
\includegraphics[height=3.2cm,width=7.4cm]{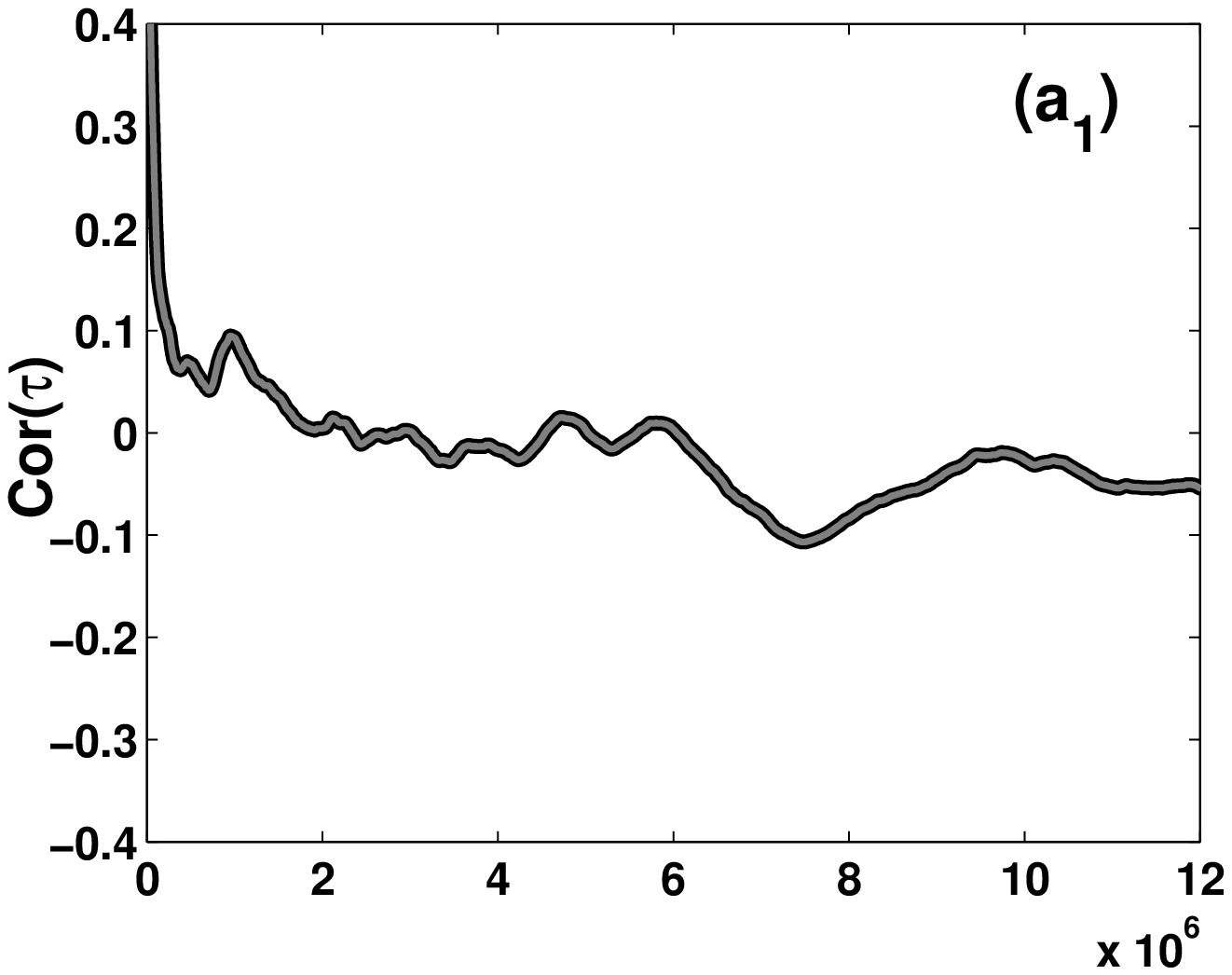}
\includegraphics[height=3.2cm,width=7.4cm]{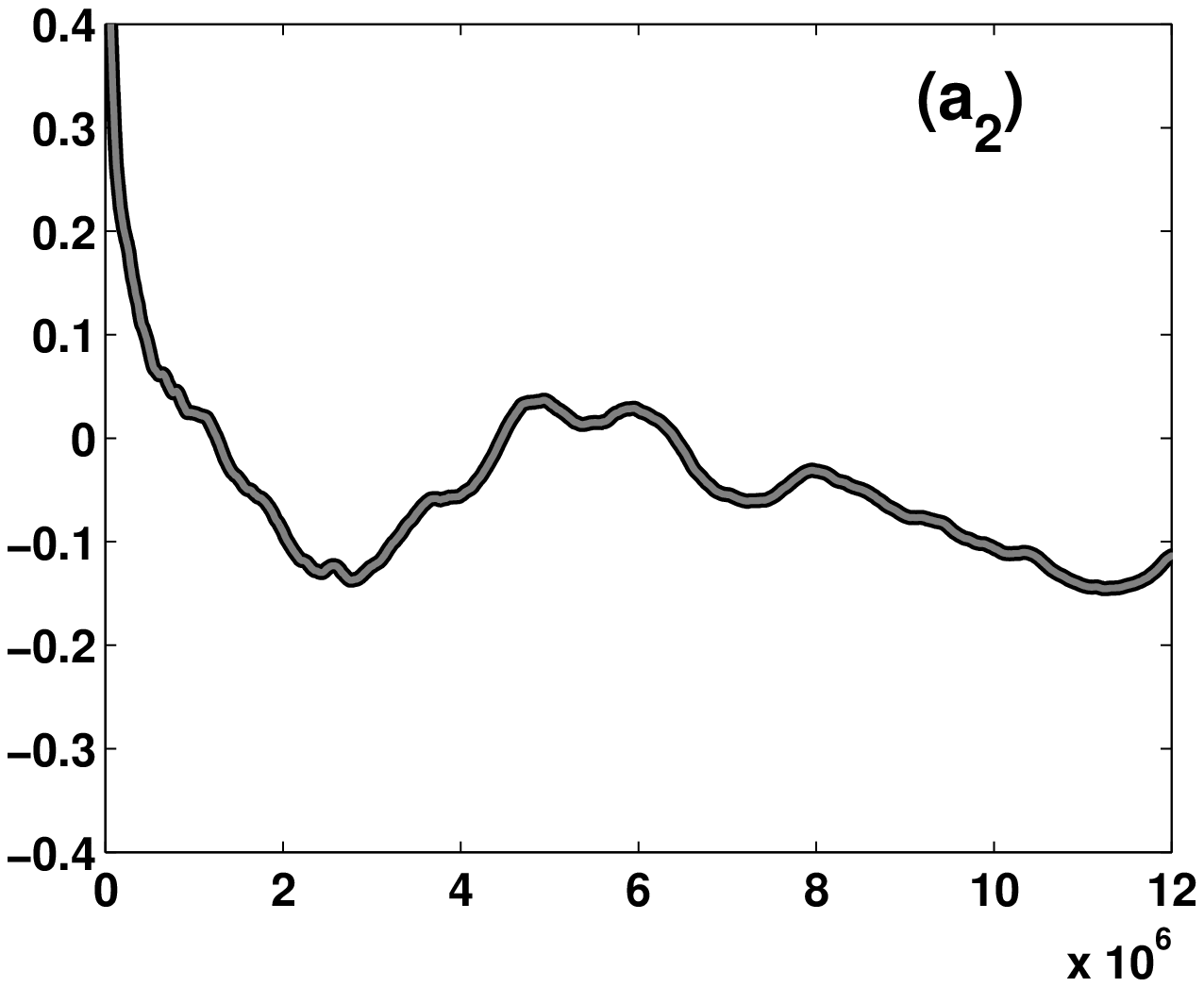}\\
\includegraphics[height=3.2cm,width=7.4cm]{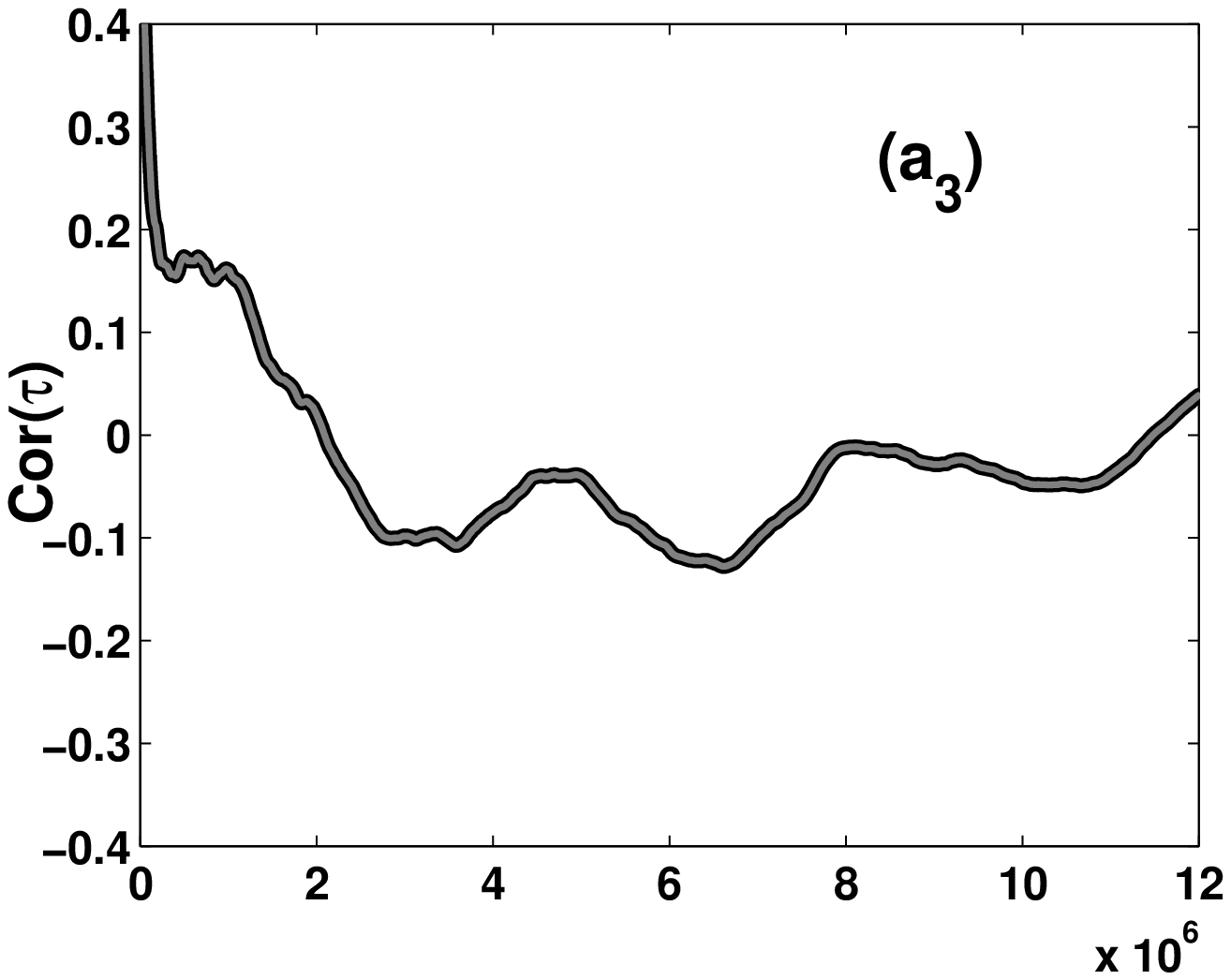}
\includegraphics[height=3.2cm,width=7.4cm]{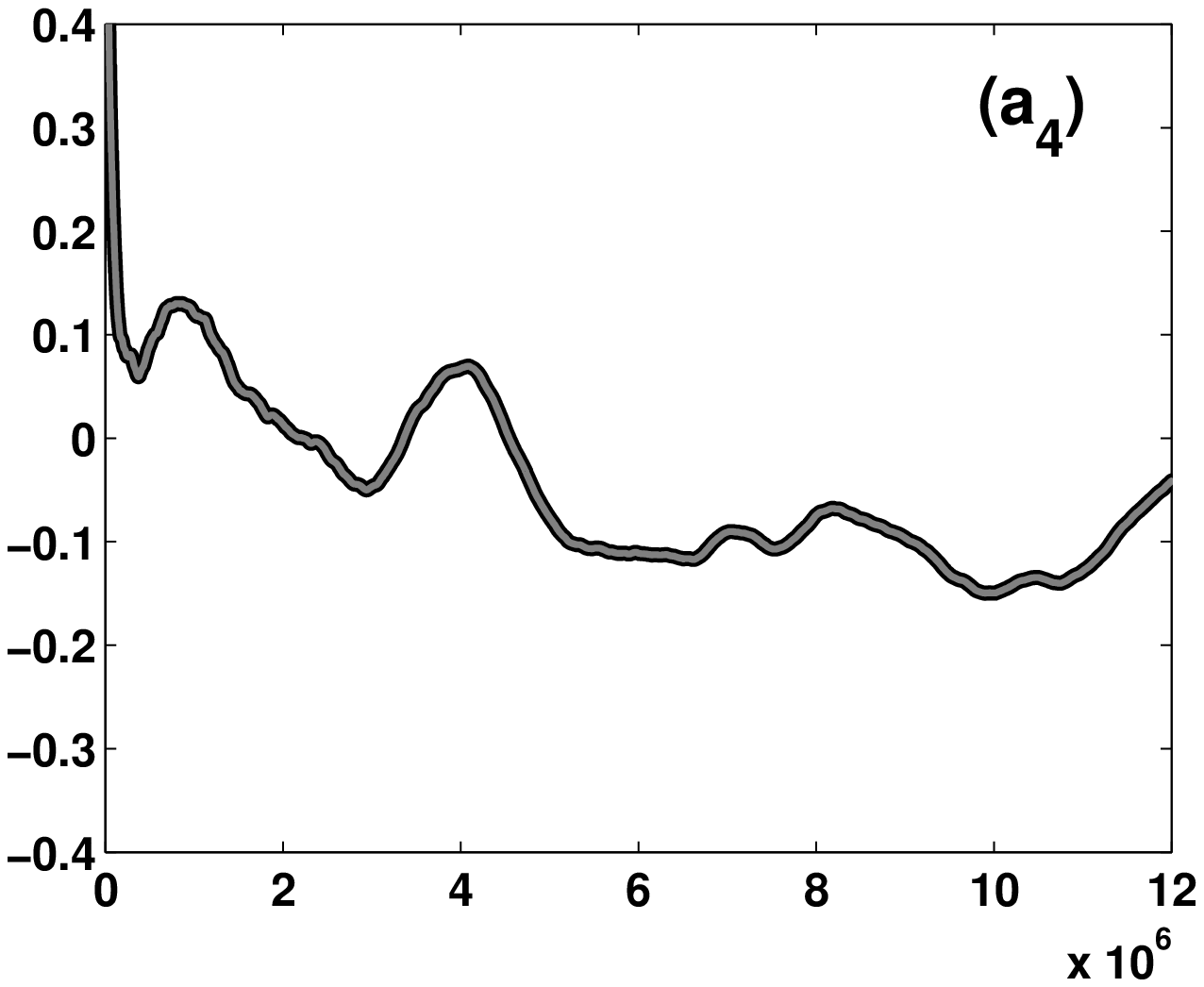}\\
\includegraphics[height=3.2cm,width=7.4cm]{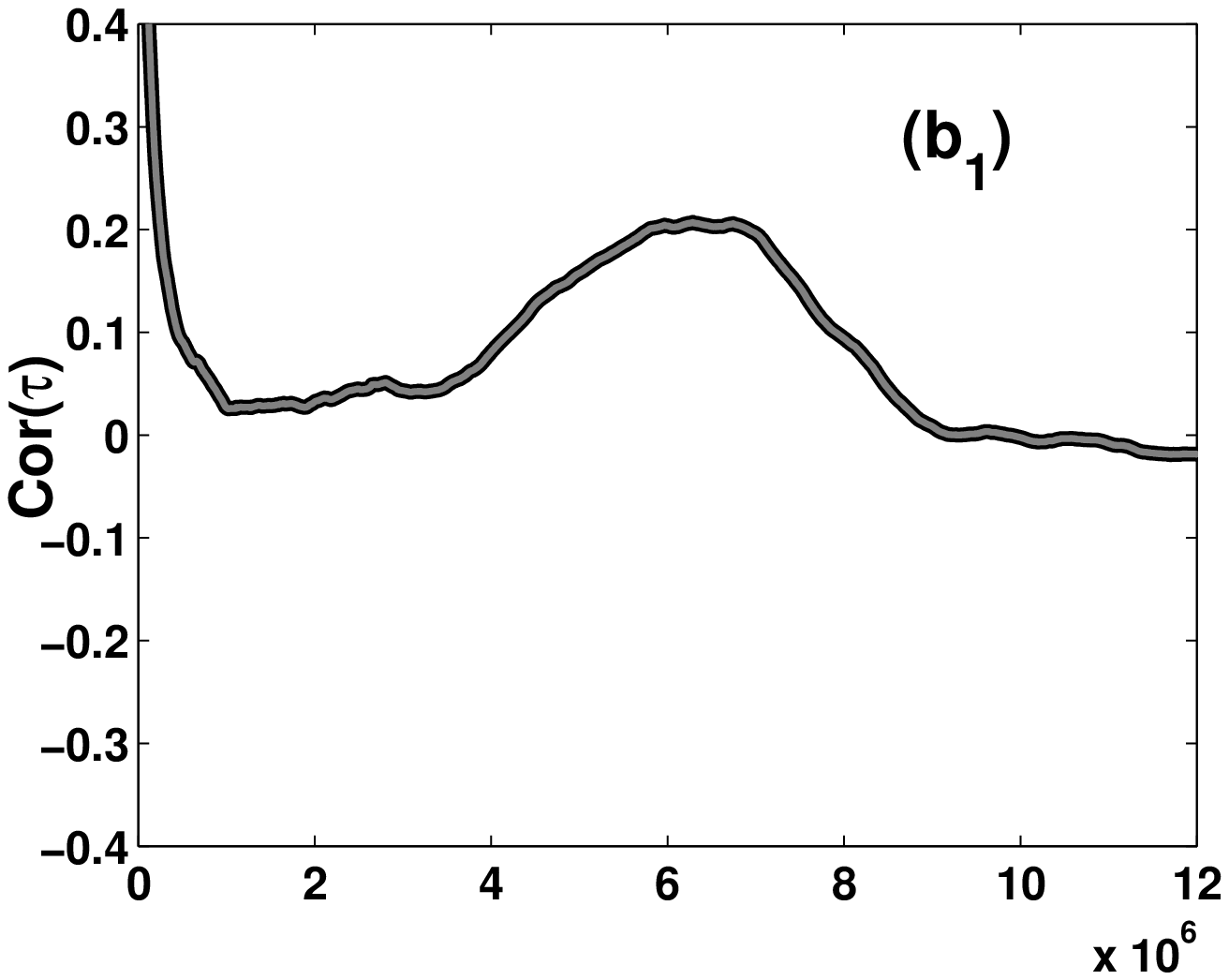}
\includegraphics[height=3.2cm,width=7.4cm]{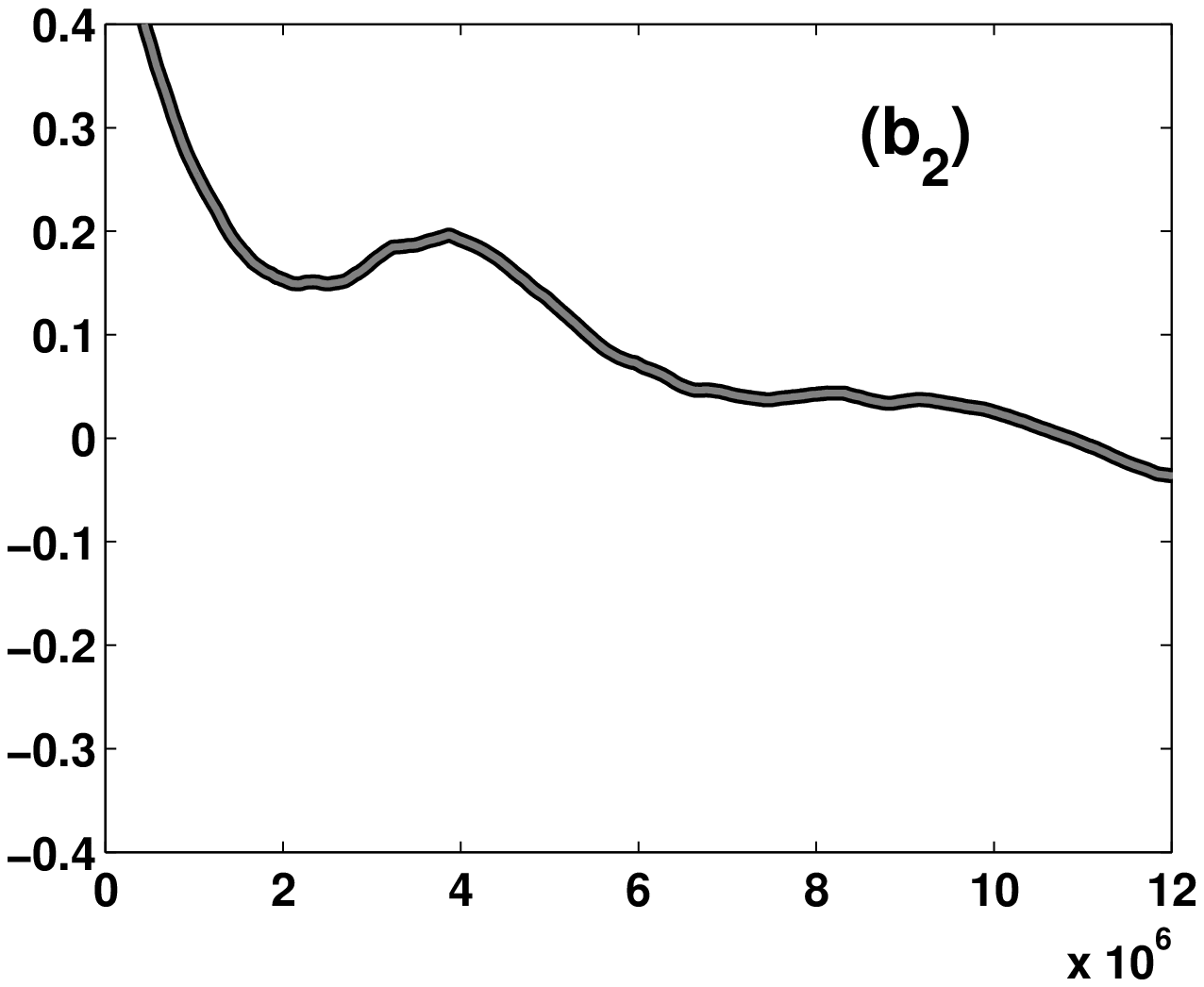}\\
\includegraphics[height=3.2cm,width=7.4cm]{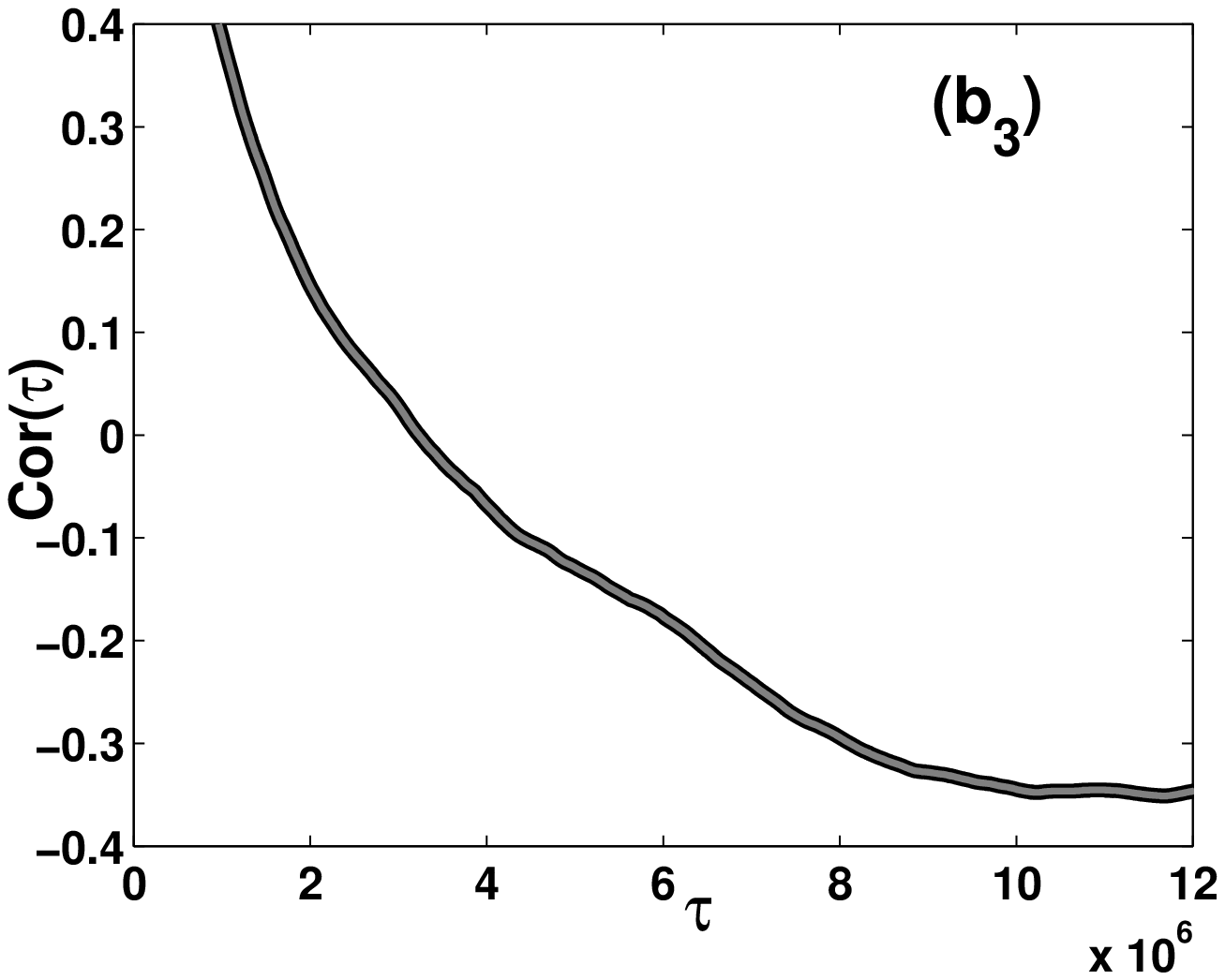}
\includegraphics[height=3.2cm,width=7.4cm]{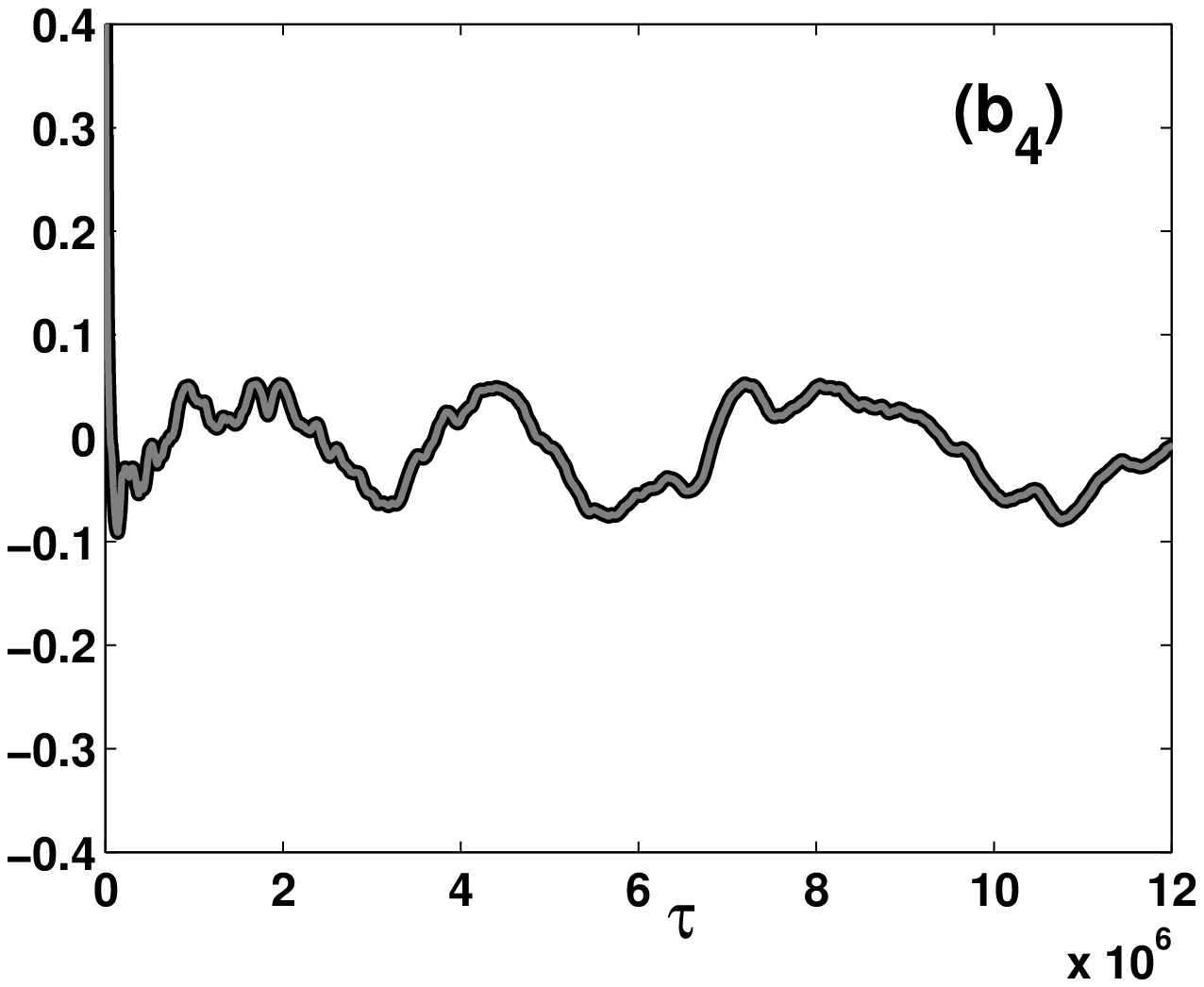}
\caption{\it
The autocorrelation function $Cor (\rho)$ as a function of the normalized time  for different values of $\gamma$ in the two transitions: Black line corresponds to the transition $A_1 \Longrightarrow A_3$, and the gray line corresponds to the transition $A_3 \Longrightarrow A_1$. The parameters are as in Table \ref{Tab1} for $S_1 (a_i)$ and $S_2 (b_i)$.
The other parameters are: $\mu=0.01$, $b=0.0$, $E=0$.
 $\gamma=0.5$ for $a_1$ and $b_1$, $\gamma=1$ for $a_2$ and $b_2$, $\gamma=1.5$ for $a_3$ and $b_3$ and $\gamma=1.7$ for $a_4$ and $b_4$.
 $a_1$ and $b_1$ correspond to $S_1$ and $a_2, b_2$ correspond to $S_2$.}
\label{fig3}
\end{center}
\end{figure}

In this Section, we investigate the possible appearance of CR,
 using the auto-correlation function (ACF) and the power spectral density tools, described in the previous Section.
Figs.\ref{fig3}($a_i-b_i$) show the variation of the auto-correlation function $Cor$, Eq.(\ref{eq11}), versus the time $t$.
The figures show the L'evy noise index  $\gamma$ (for increasing values from $\gamma=0.1$ to $\gamma=1.5$) at a fixed value of the noise intensity $D=0.1$.
The behavior of $Cor(\rho)$ together with the Gaussian case $\gamma=2$, show a strong dependence upon the initial conditions, for the transitions starting from the inner ($A_1$) or the outer ($A_3$) attractors are markedly different.

\begin{figure}
\begin{center}
\includegraphics[height=4.4cm,width=7.4cm]{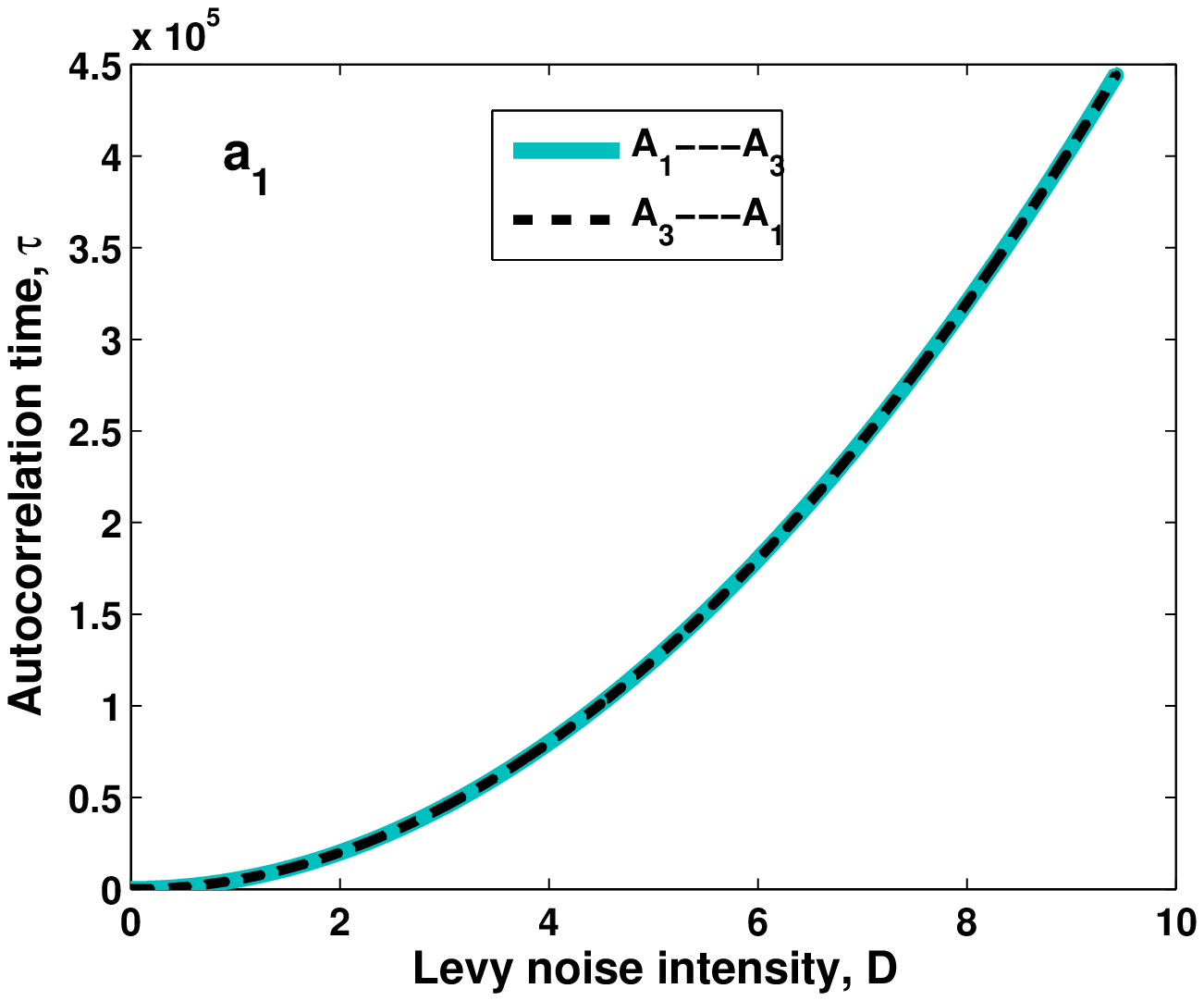}
\includegraphics[height=4.4cm,width=7.4cm]{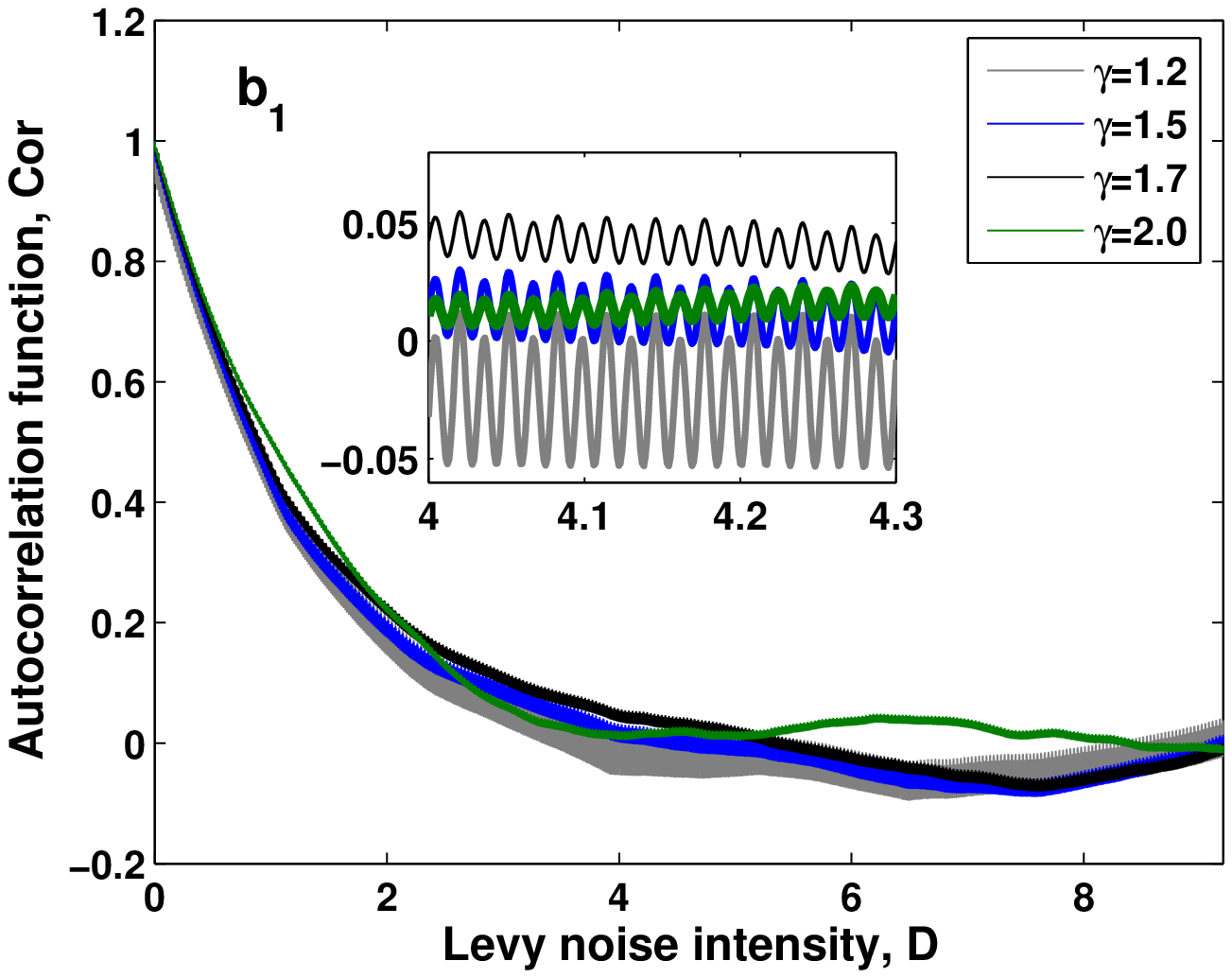}\\
\includegraphics[height=4.4cm,width=7.4cm]{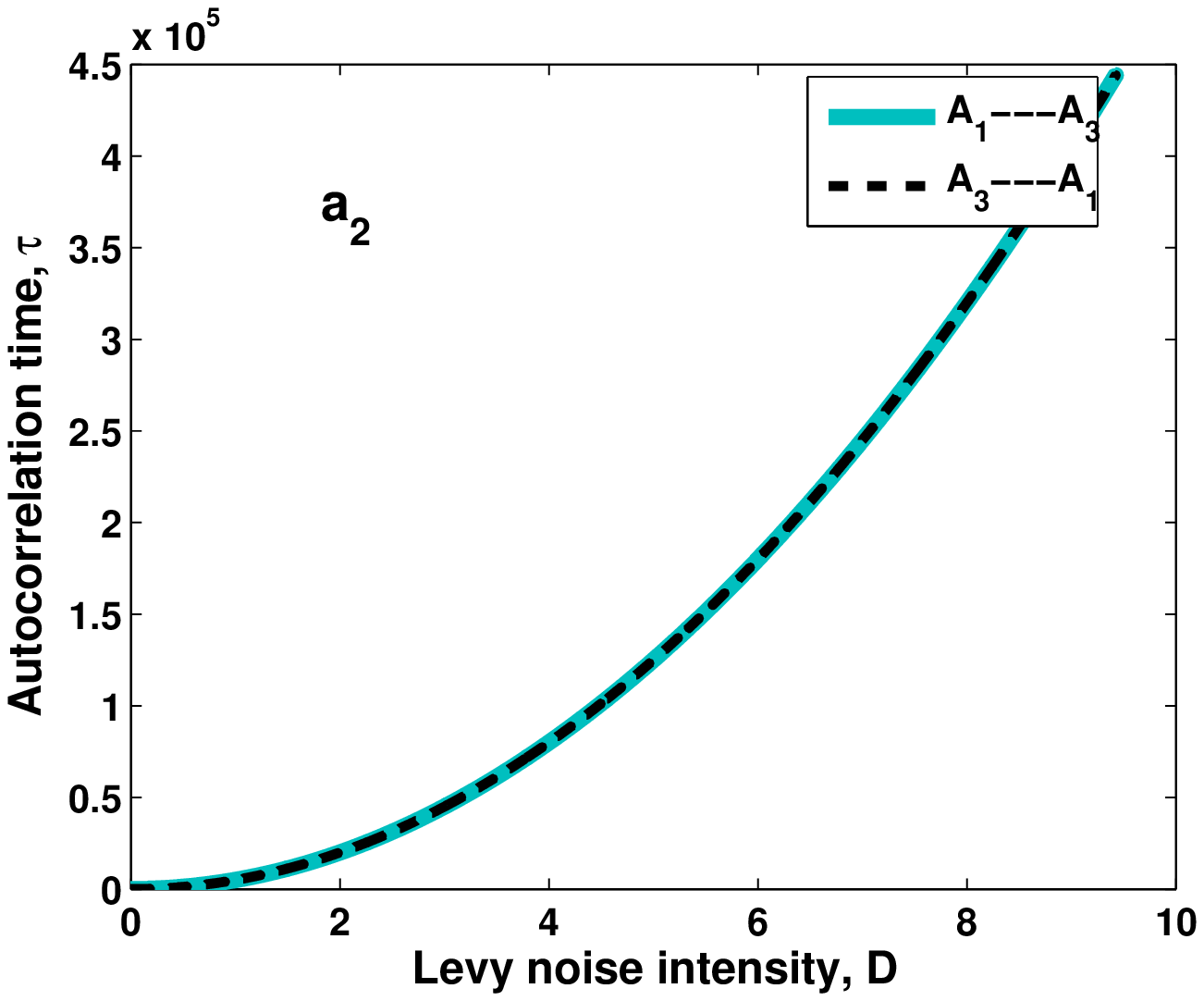}
\includegraphics[height=4.4cm,width=7.4cm]{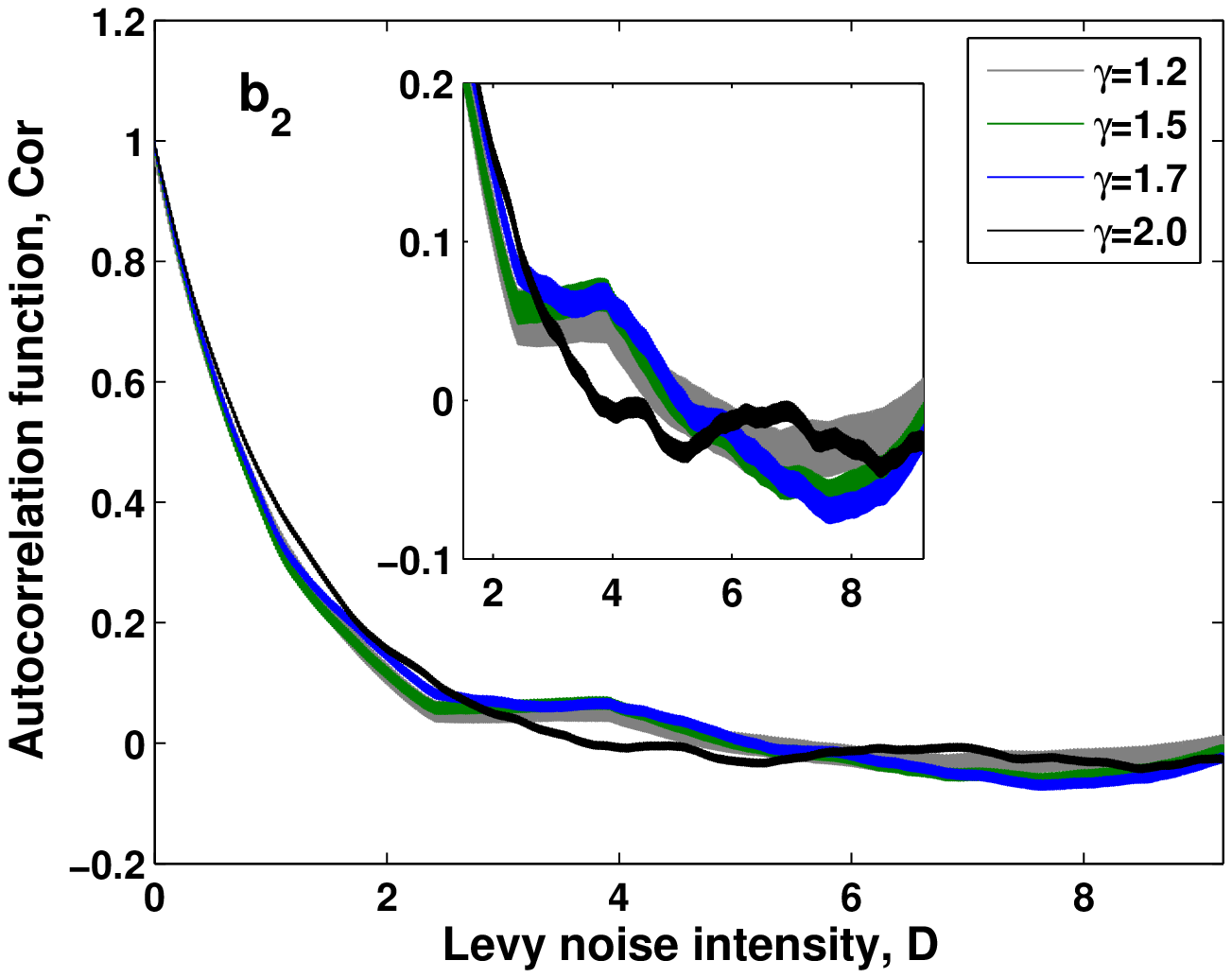}
\caption{\it
($a_i$): The autocorrelation time $\tau$ versus $D$ .
($b_i$): Variation of the autocorrelation function versus $D$ for both transitions
with $\tau= 1.5$.
The other parameters are: $S_1$ for $(a_1,b_1)$ and $S_2$
for $(a_2,b_2)$ (refer to Table \ref{Tab1}), $\mu= 0.01$, and $b= 0.0$.
}
\label{fig4}
\end{center}
\end{figure}

 Figs.\ref{fig4}($a_i$) show the dependence of the autocorrelation time $\tau$ versus $D$,
while in Fig.\ref{fig4}($b_i$) it is displayed the dependence of $Cor$ versus $D$  for several different values of  $\gamma$.
The data are collected for both transitions, $A_1\Longrightarrow A_3$ or $A_3 \Longrightarrow A_1$.
The statistical average is independent of the initial conditions.
One observes for  both transitions that, $D$ causes a steady-state response
and increases the autocorrelation time (for $\gamma=0.5$).
As expected, in  both transitions the autocorrelation time $\tau$, defined in Eq.(\ref{eq13}), increases with the noise intensity $D$, while the ACF $Cor$ decreases.
For  both transitions, the behavior is  qualitatively the same, but quantitatively changes if the control parameters are changed (see for example Fig.\ref{fig4}($b_i$).

\begin{figure}
\begin{center}
\includegraphics[height=4.4cm,width=7.4cm]{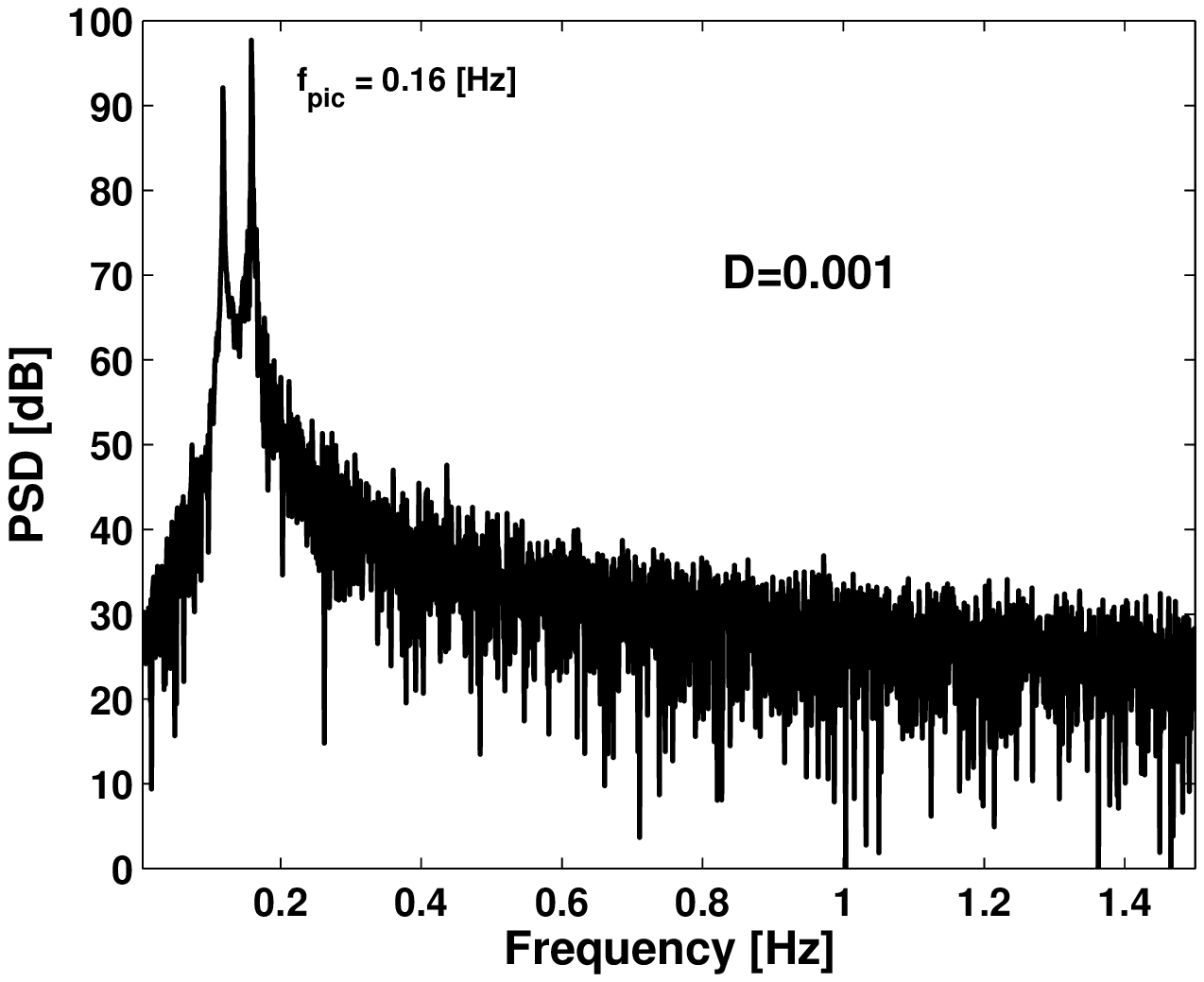}
\includegraphics[height=4.4cm,width=7.4cm]{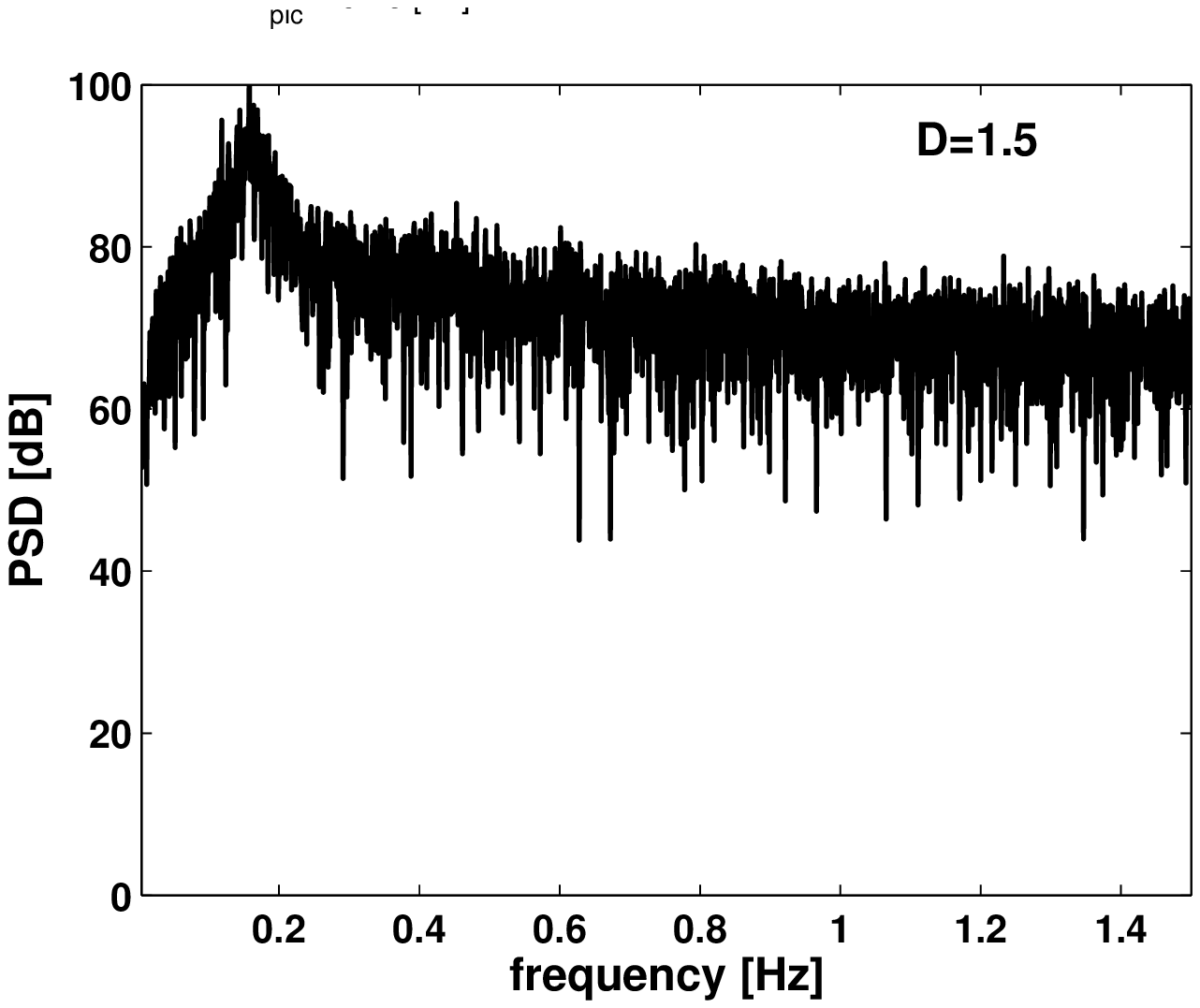}\\
\includegraphics[height=4.4cm,width=7.4cm]{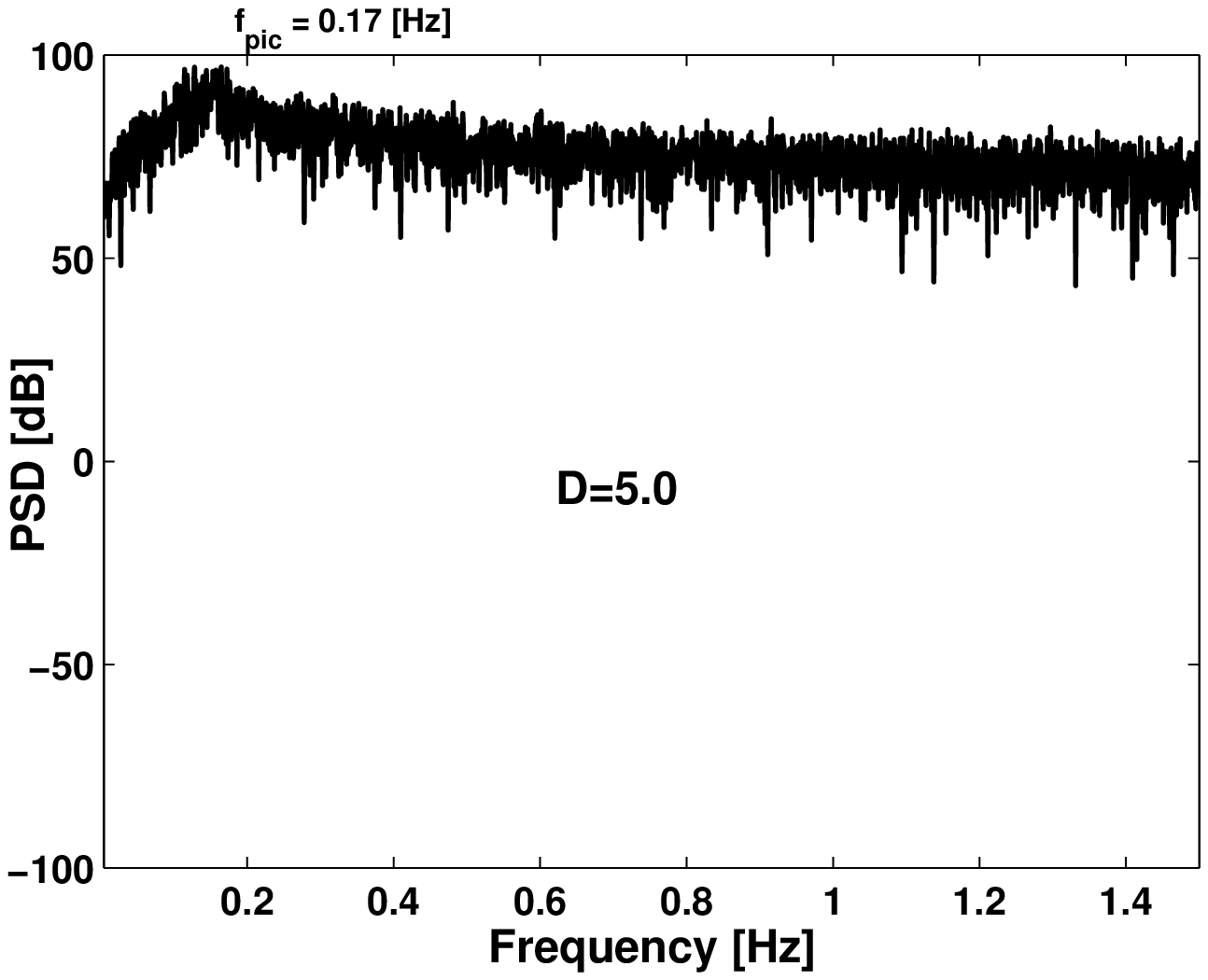}
\includegraphics[height=4.4cm,width=7.4cm]{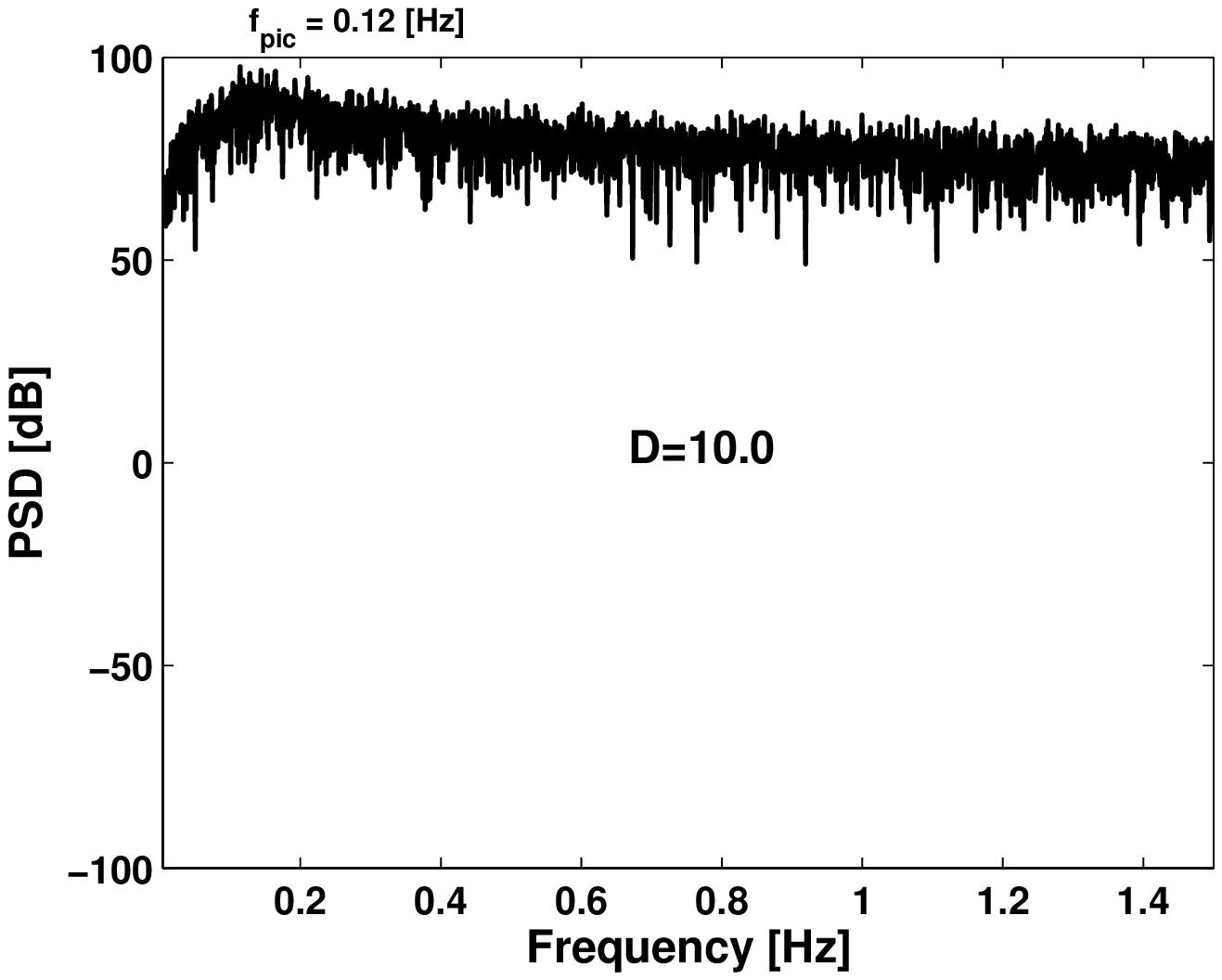}\\
\caption{\it
 Power spectral density in the presence of a periodic external force
  and LN with four values of  $D$.
 The parameters of the system are $\alpha=0.114$,
 $\beta=0.003$, $E=0.5$, $\gamma=1.5$, $\mu=0.01$, $\omega={\sqrt{5.0}}/{3}$.}
\label{fig5}
\end{center}
\end{figure}

\begin{figure}
\begin{center}
\includegraphics[height=4.4cm,width=7.4cm]{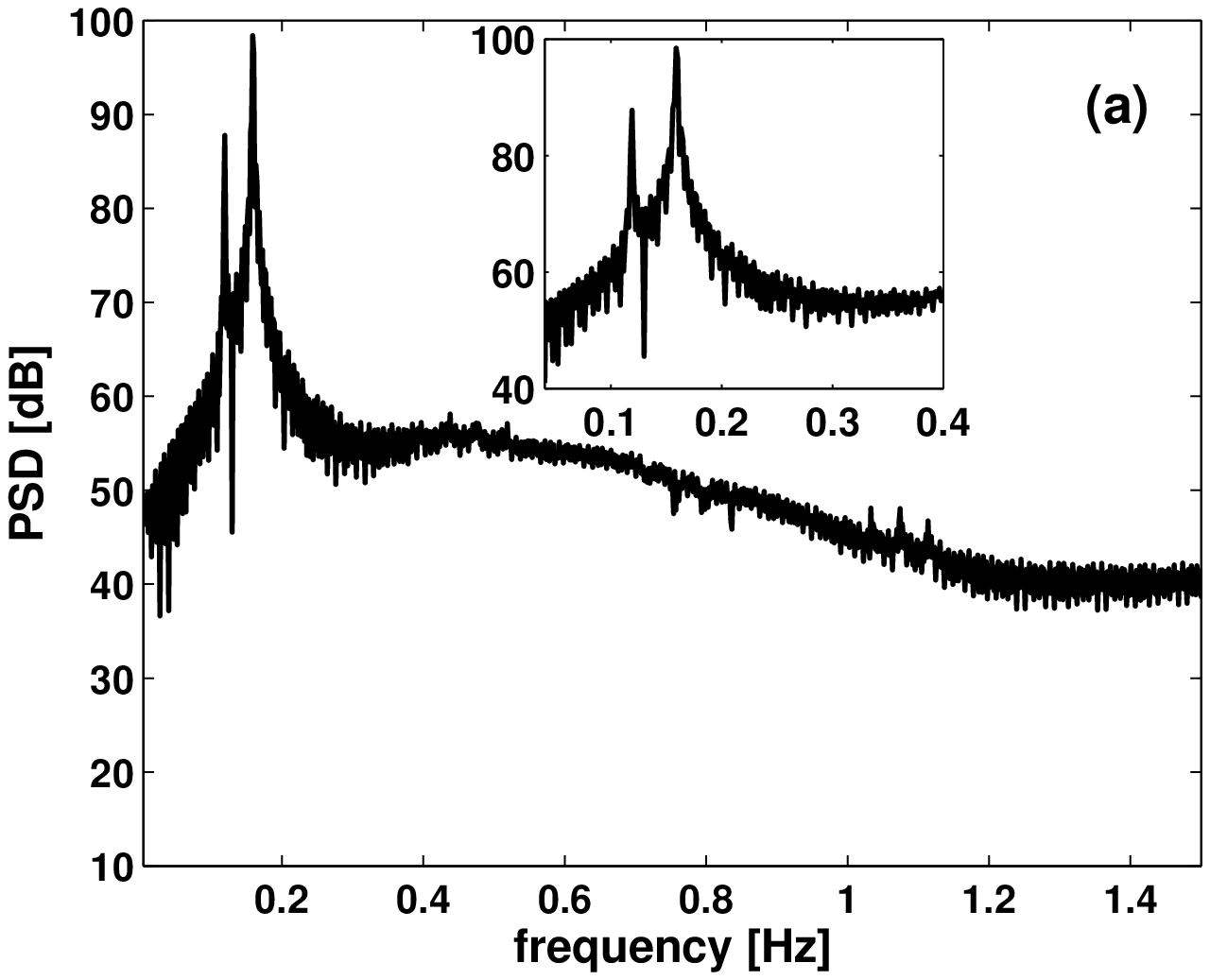}
\includegraphics[height=4.4cm,width=7.4cm]{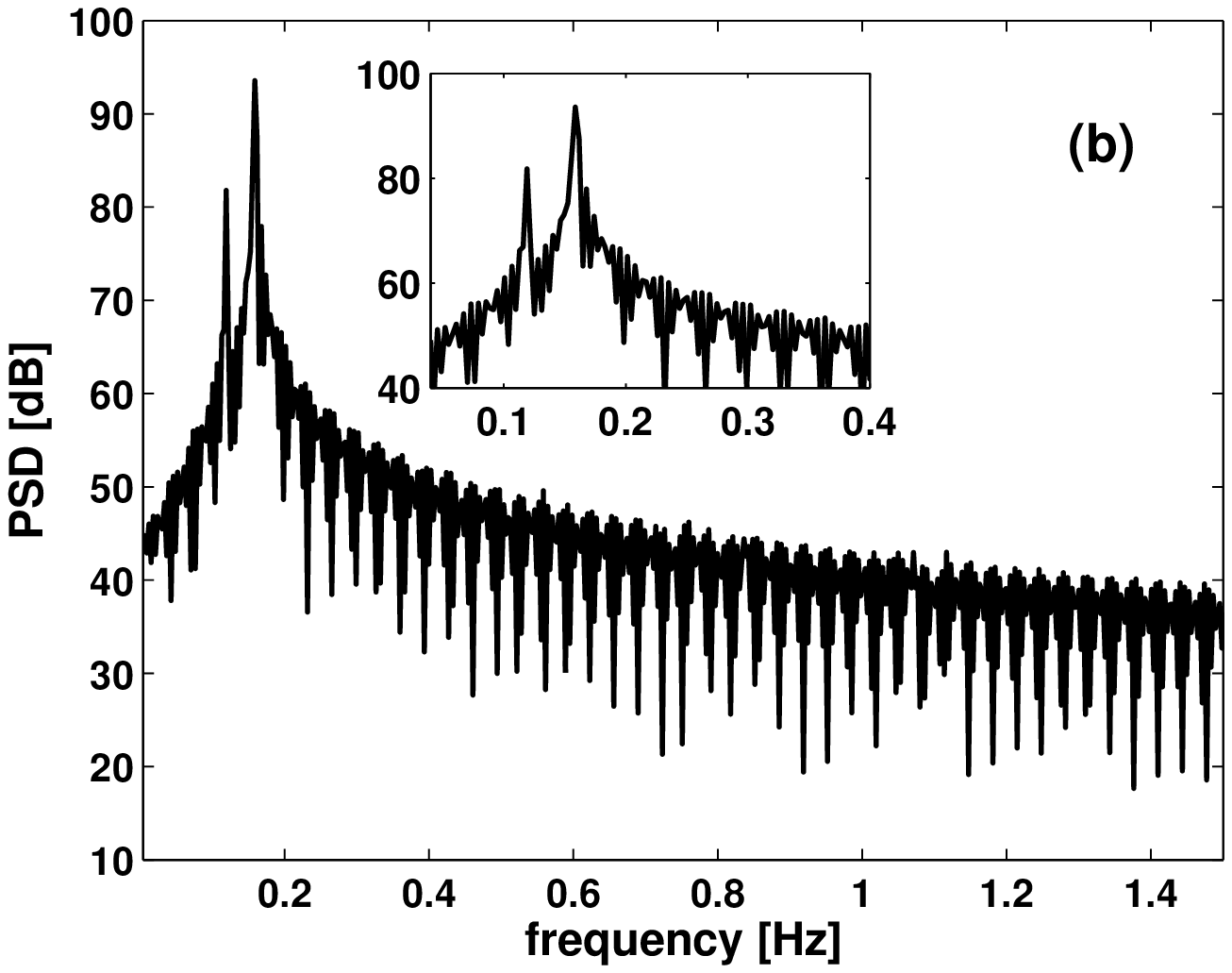}\\
\includegraphics[height=4.4cm,width=7.4cm]{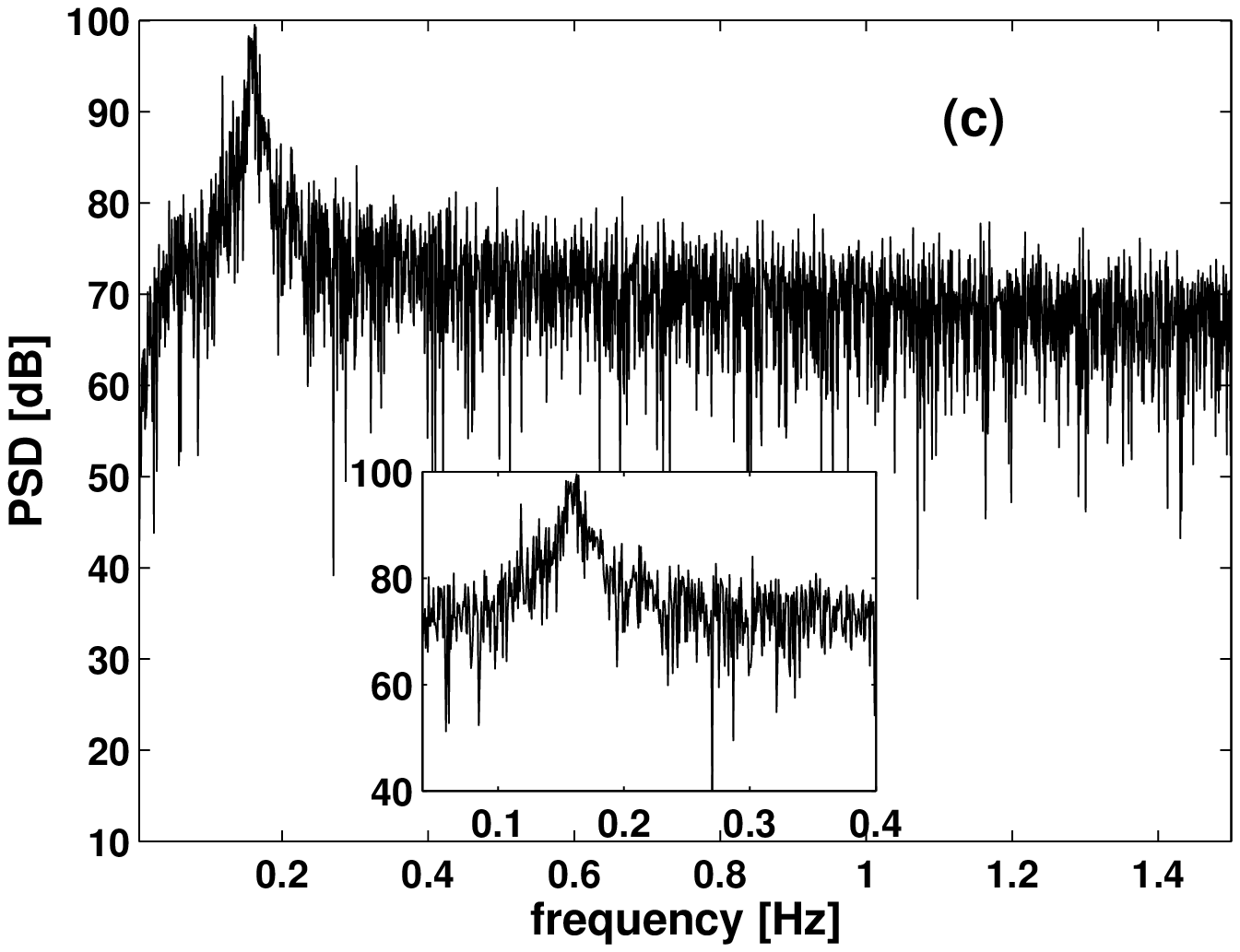}
\includegraphics[height=4.4cm,width=7.4cm]{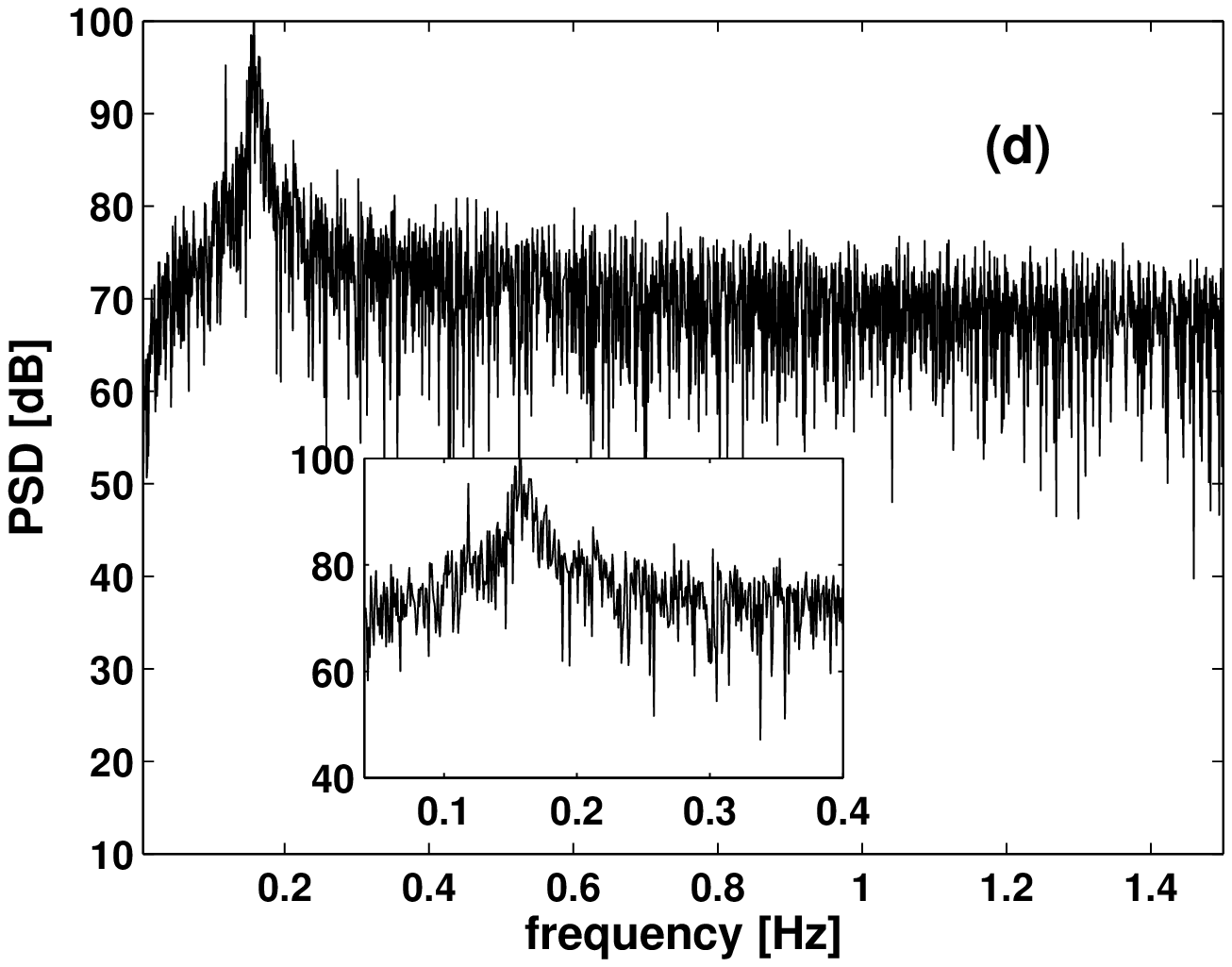}\\
\includegraphics[height=4.4cm,width=7.4cm]{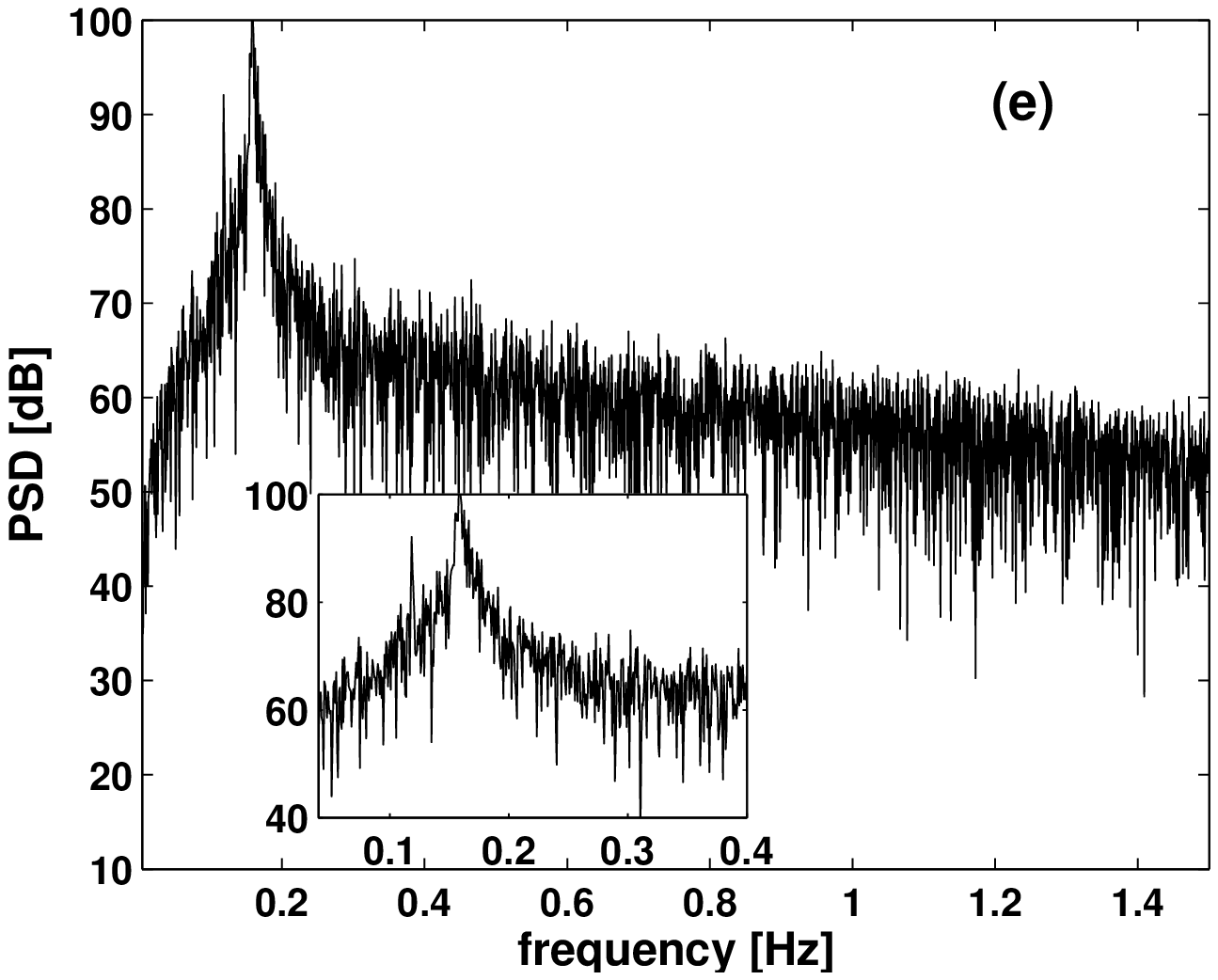}
\includegraphics[height=4.4cm,width=7.4cm]{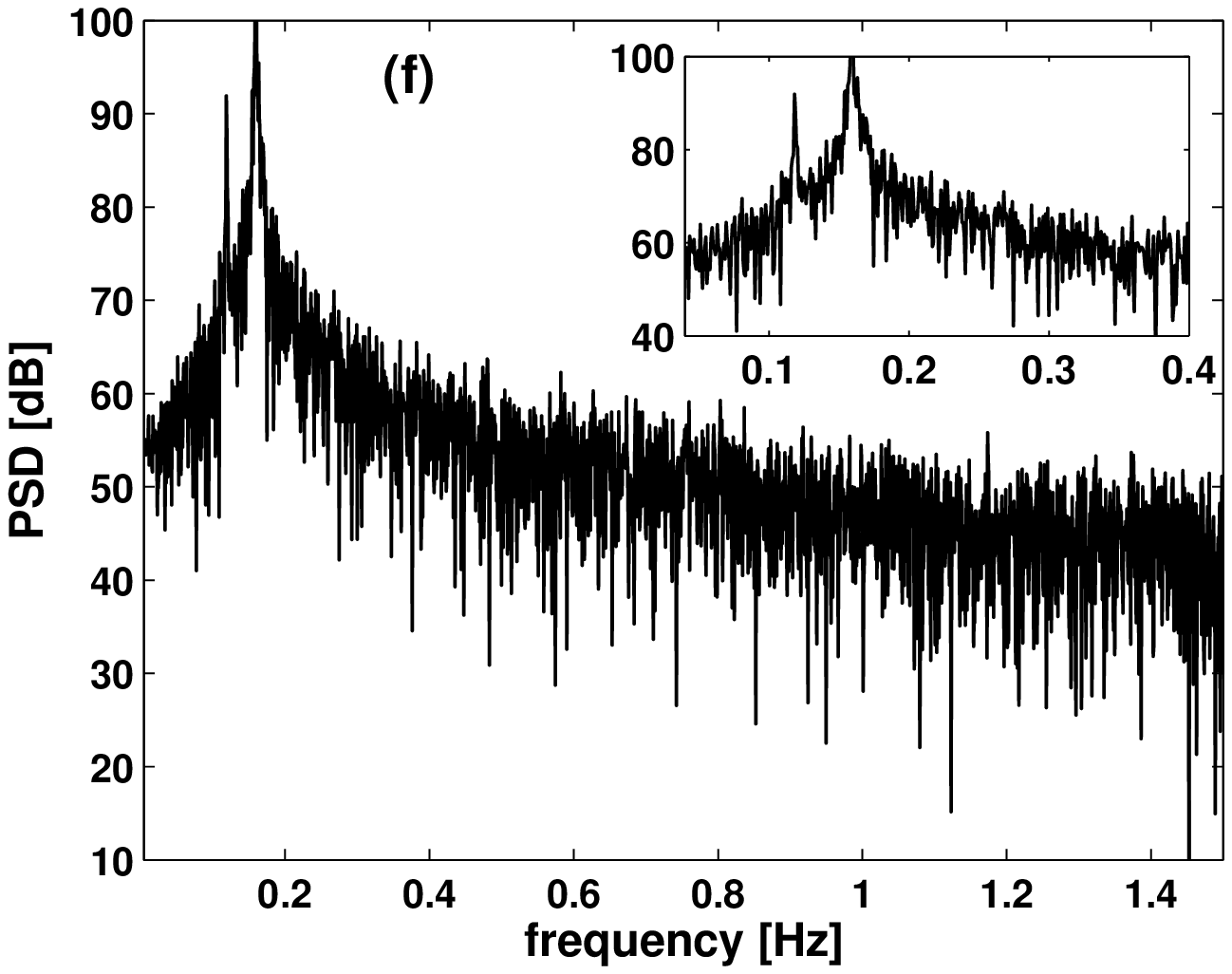}
\caption{\it
 Effect of $\gamma$ on the variation of  PSD in the presence of LN
  excitation, periodic external force for four values of
  $\gamma$ ((a): $\gamma=0.15$, (b): $\gamma=0.45$, (c): $\gamma=0.8$, (d): $\gamma=1.0$, (e):
 $\gamma=1.5$, (f): $\gamma=2.0$). The parameters of the system are
 $\alpha=0.114$, $\beta=0.003$, $E={1}/{2}$, $D=0.01$, $\mu=0.01$, $\omega={\sqrt{5.0}}/{3}$.}
\label{fig6}
\end{center}
\end{figure}

\begin{figure}
\begin{center}
\includegraphics[height=4.4cm,width=7.4cm]{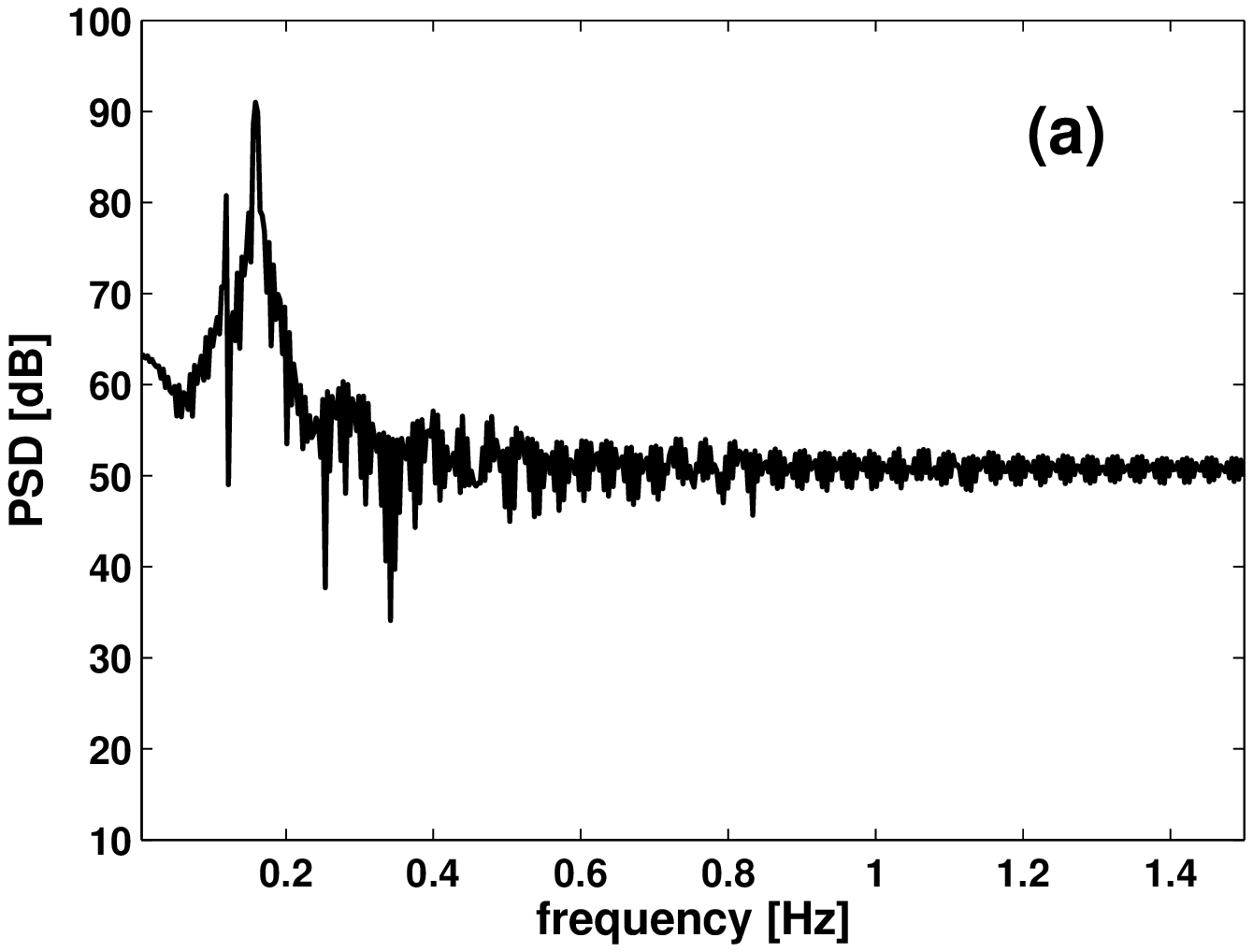}
\includegraphics[height=4.4cm,width=7.4cm]{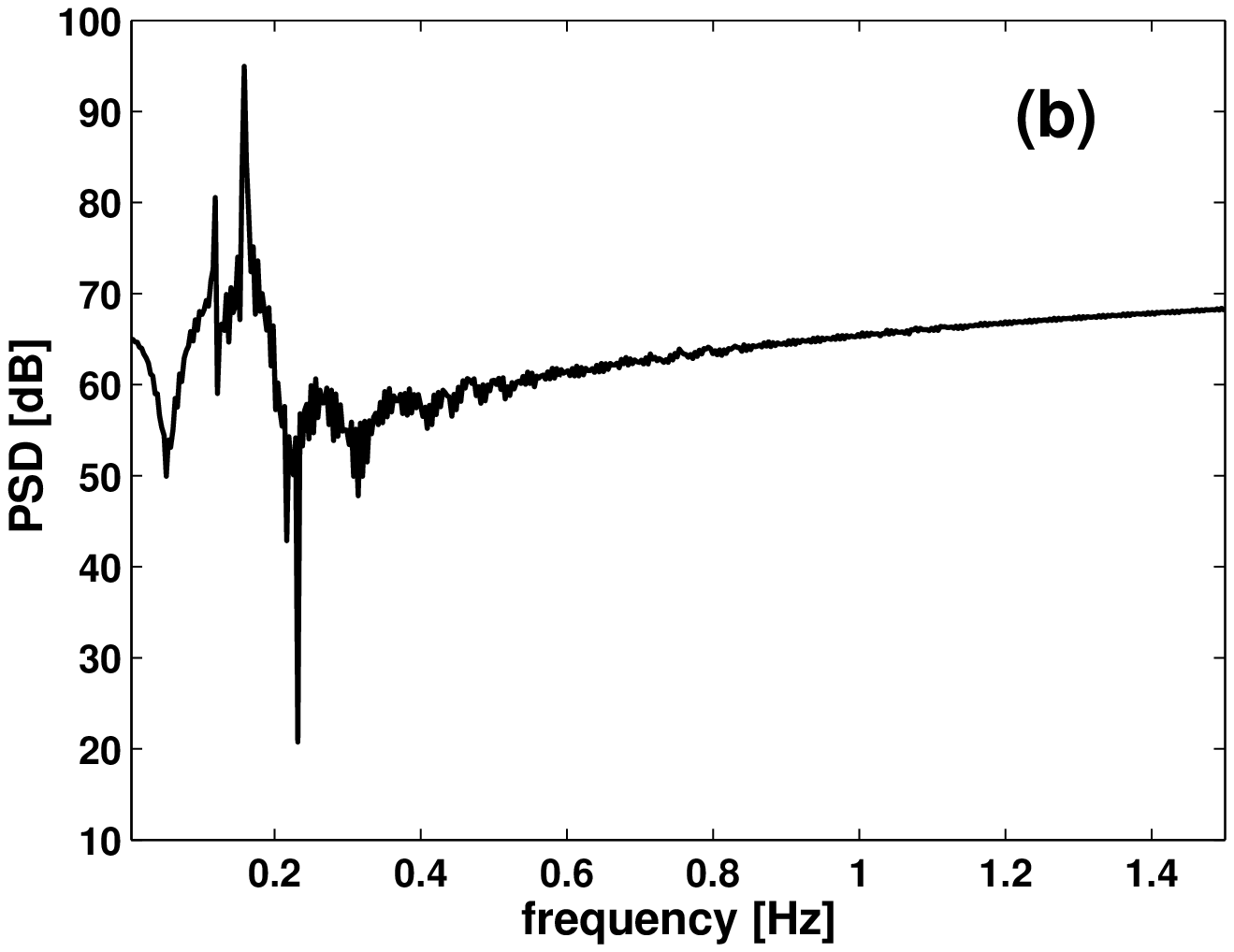}\\
\includegraphics[height=4.4cm,width=7.4cm]{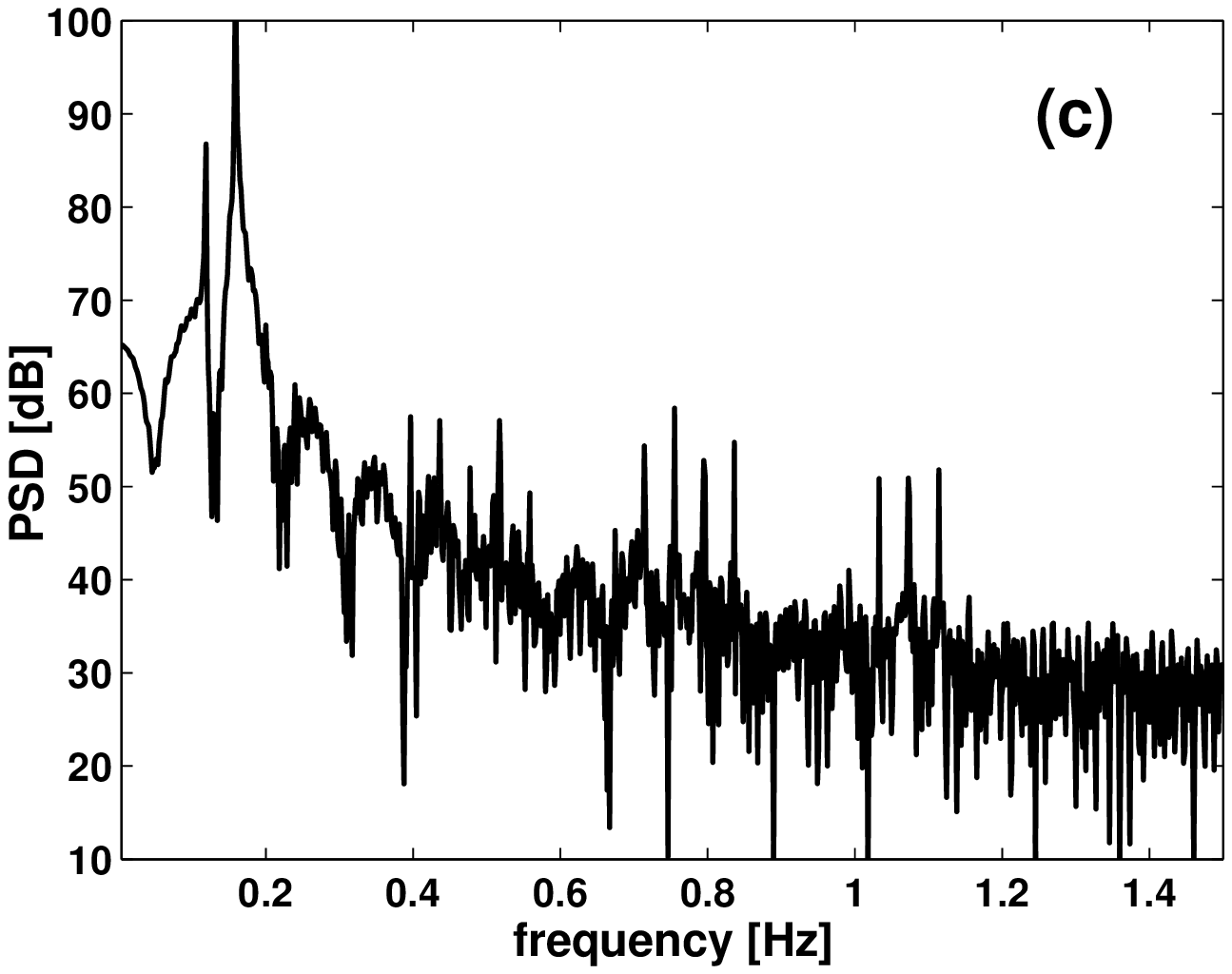}
\includegraphics[height=4.4cm,width=7.4cm]{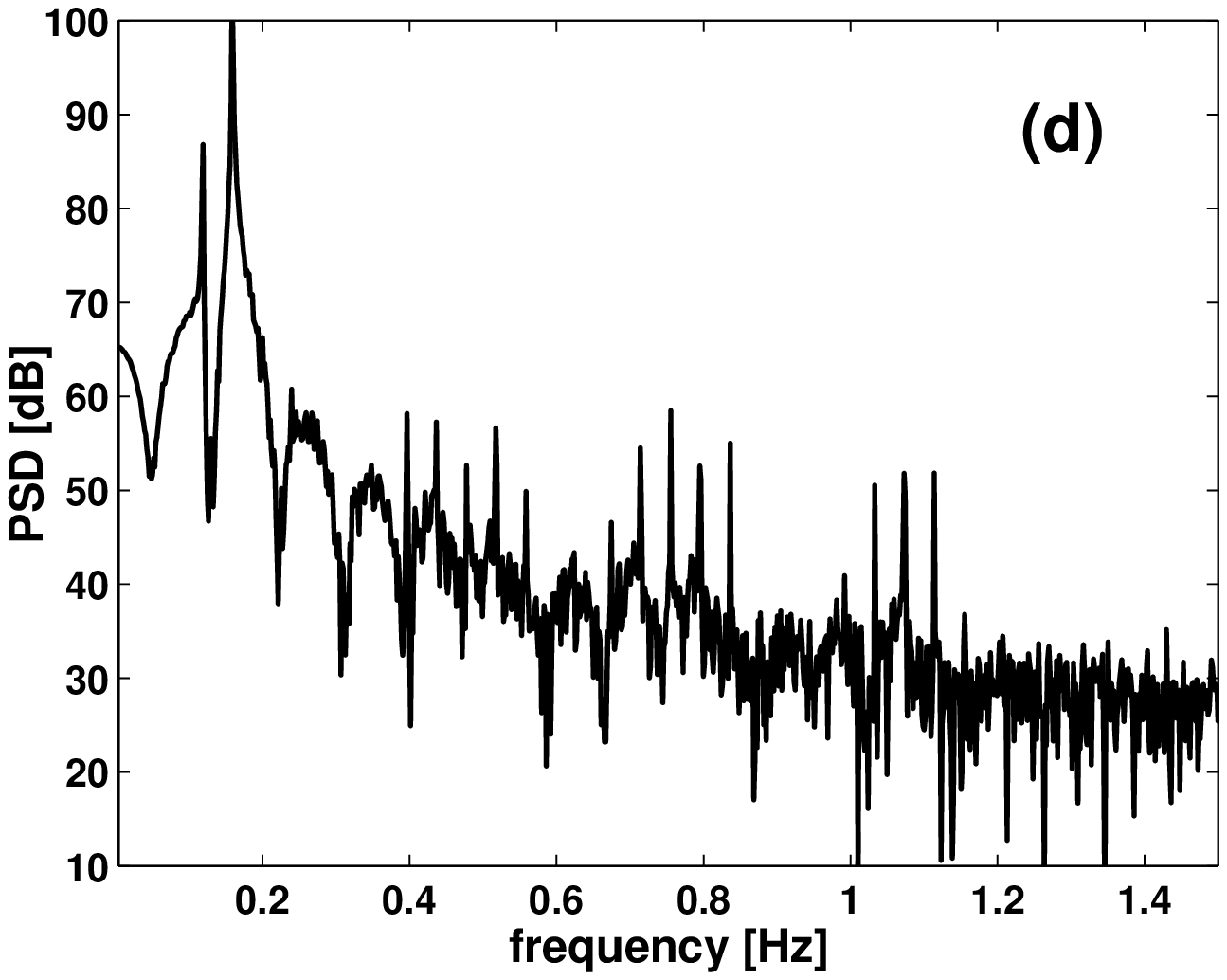}
\caption{\it
 Effects of the LN parameter $\gamma$ on the PSD in the presence non-Gaussian noise and periodic external forcing for $\gamma=0.45$ (a), $\gamma=0.8$
   (b), $\gamma=1.3$ (c), and $\gamma=1.9$ (d). The  initial
conditions are chosen on the  attractor  $A_3$ and the control parameters in
the birhythmic region:
    $\alpha=0.114$, $\beta=0.003$, $E={1}/{2}$, $D=0.01$, $\mu=0.01$, $\omega={\sqrt{5}}/{3}$.}
\label{fig6a}
\end{center}
\end{figure}

The PSDs of the van der Pol system (\ref{eq1}) for few values of the noise intensity
 $D$ are displayed in Fig.\ref{fig5}.
 The system is initialized on  the inner attractor, and therefore the power spectrum refers to fluctuations around the orbits $A_1$.
 It is  evident that there is a qualitative effect of the noise on the peaks of the spectrum.
In Fig.\ref{fig6} the PSD is evaluated for the other attractor, $A_3$.
It is evident that the behavior of PSD strongly depends also on the L\`evy distribution index $\gamma$.
Increasing the L\`evy index $\gamma$ from $0.15$, one observes a small peak around a frequency $\Omega=0.14$, while the main peak stays around the  frequency $\Omega=0.2$.
This main peak increases if the index $\gamma$ is raised up to $\gamma=0.45$, while the lower peak disappears.
The  amplitude of one of the harmonic first increases to reach the maximum and then decreases.
The peak of the PSD at the modulation frequencies is not suppressed as $\gamma$  of the  L\`evy noise spectrum becomes large enough and the signal amplitude diminishes for various values of $\gamma$.
In all figures, the PSD exhibits a sharp peak at the  frequency $\omega$, which indicates that there is a stronger signal  at frequency $\omega=0.2$.
The value of this peak  is significant to indicate the possibility of the occurrence of SR, implying that noise can enhance the output power spectrum of the system at some special frequency, and thus resembles stochastic  resonance, that is an optimal output with  a  proper   adjustment  of the noise amplitude $D$.
For the attractor $A_3$ (see for example Fig.\ref{fig6}), the peaks   appear in general less evident.
In the two figures, it seem that the behavior of PSD depends on the central frequency, on the nearby   attractor, and the portion of the PSD under consideration.
 Figure \ref{fig6a} shows the effects of the LN parameter $\gamma$ on 
 the PSD in the presence non-Gaussian noise and periodic external forcing, 
 for the escape transition $A_3\Longrightarrow A_1.$

The Figures demonstrate that SR in this system has certain connections    with the nonconventional SR, although only a very rough match between   the driving frequency and the noise-tuned frequency peak has been observed.
 The nonconventional SR in this under damped bistable van der  Pol oscillator \cite{Mbakob3} should reflect the complexity  of the    pseudo-potential and this complexity has been  reported from the     viewpoint of the dynamical states
  identified from the phase difference of the trajectory  of random moving particle \cite{Mbakob1,Mbakob2,Mbakob3,Mbakob4}, as will be examined in the next Section.

\section{Stochastic Resonance detection}
\label{stochastic}

In this Section the SNR given by Eqs. (\ref{eq14},\ref{eq15}) is employed to quantify the occurrence of SR in the birhythmic system (\ref{eq1}).
The occurrence of SR may be detected through the presence of a peculiar nonlinear behavior of the response of bistable systems exposed to increasing intensity of the noise level.
In fact, at low noise level, the particle oscillates at the bottom of one of the potential wells for a long time, and rarely switches to the other potential wells.
However, if the noise intensity is increased, the frequency of the switches grows and can become coherent with some natural period of the system.
As noise is further increased, the switches from an attractor to another become purely random, and thus the coherence is eventually lost.

\subsection{Stochastic resonance under Gaussian noise}
\label{stochastic_gaussian}

In this Section, a numerical method of moment based on non perturbation expansion is presented to observe SR under Gaussian noise (that is equivalent to L\`evy noise with $\gamma=2.0$).
Therefore, our aim here is to make a comparison of the  two numerical methods to make sure that the results are consistent with Refs. \cite{Chamgoue}.
The results summarized in Table \ref{Tab2} confirm that the numerically generated GN and LN for $\gamma=2$ hold the same results.

\begin{table}
\begin{center}
\begin{tabular}{|l|c|c|c|c|}
\hline
$S_i=(\alpha,\beta)$&  $D_{SR}$ for GN& $SNR_{max}$ for GN & $D_{SR}$ under LN& $SNR_{max}$ under LN\\
\hline

$(0.12; 0.0032)$           & $0.4138$    & $2.8752$ & $0.4063$ & $2.8747$\\
\hline \hline
$(0.1476; 0.0053)$           & $4.2347$    & $2.6891$ & $4.2748$ & $2.6808$\\
\hline
$(0.0675; 0.0009)$           & nothing    & nothing & nothing &nothing\\
\hline
$(0.1547; 0.006)$           & $4.6874$    & $2.7799$ & $4.6881$ & $2.7802$\\
\hline
$(0.145; 0.005)$           & $2.3043$    & $2.8260$ & $2.3202$ & $2.7876$\\
\hline
\end{tabular}
\caption{Comparison between the results obtained for the  birhythmic system under GN and  LN with $\gamma=2.0$. The results show that the numerical method for employed to generate LN is consistent with a straightforward generation of Gaussian noise.}
\end{center}
\label{Tab2}
\end{table}

\subsection{Stochastic resonance under  L\`evy noise}

\begin{figure}
\begin{center}
\includegraphics[height=6.4cm,width=7.4cm]{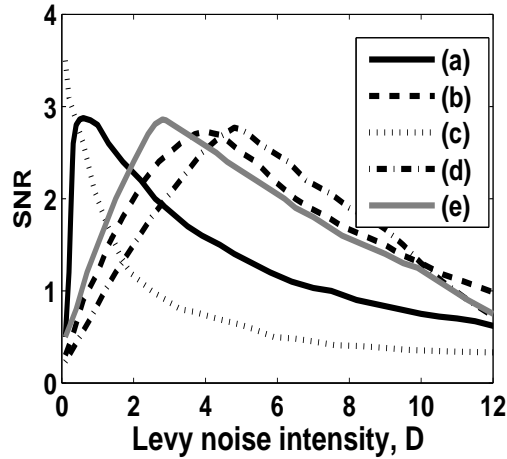}
\caption{\it
The SNR versus noise intensity $D$ for few control parameters.
The data correspond to the dynamical states for  initial
conditions chosen on the inner attractor  $A_1$ and the control parameters in
the birhythmic region:
$(a)$: $\alpha = 0.12$, $\beta =0.0032$;
$(b)$: $\alpha=0.1476$, $\beta= 0.0053$;
$(c)$: $\alpha=0.0675$, $\beta=0.0009$;
$(d)$: $\alpha=0.1547$, $\beta= 0.006$;
$(e)$: $(\alpha=0.145$, $\beta= 0.005$.
All data refer to $\mu=0.001$, $E=0.1$, $\omega=0.2$.}
\label{fig6b}
\end{center}
\end{figure}

Stochastic resonance  induced by LN (\emph{i.e.},  $\gamma<2$) in  the  system (\ref{eq1})  is  investigated.
The role of the L\`evy index $\gamma$  on SR is highlighted through the analysis of the PSD, to retrieve the SNR per Eq. (\ref{eq15}), as illustrated in Sect. \ref{numerical}.
The external drive frequency $\omega$ is fixed to focus the role of the orbits' period and of the symmetry of the pseudo-potential.
The maximum in the SNR, detected adjusting the noise intensity, is the signature of the presence of SR.
Figure \ref{fig6b} shows the effects of the control parameters  on the SNR dependence of the noise intensity. It found
 that the stochastic resonance doesn't appeared with the following control parameters: $(\alpha,\beta)=(0.0675,0.0009)$.

\begin{figure}
\begin{center}
\includegraphics[height=5.4cm,width=7.4cm]{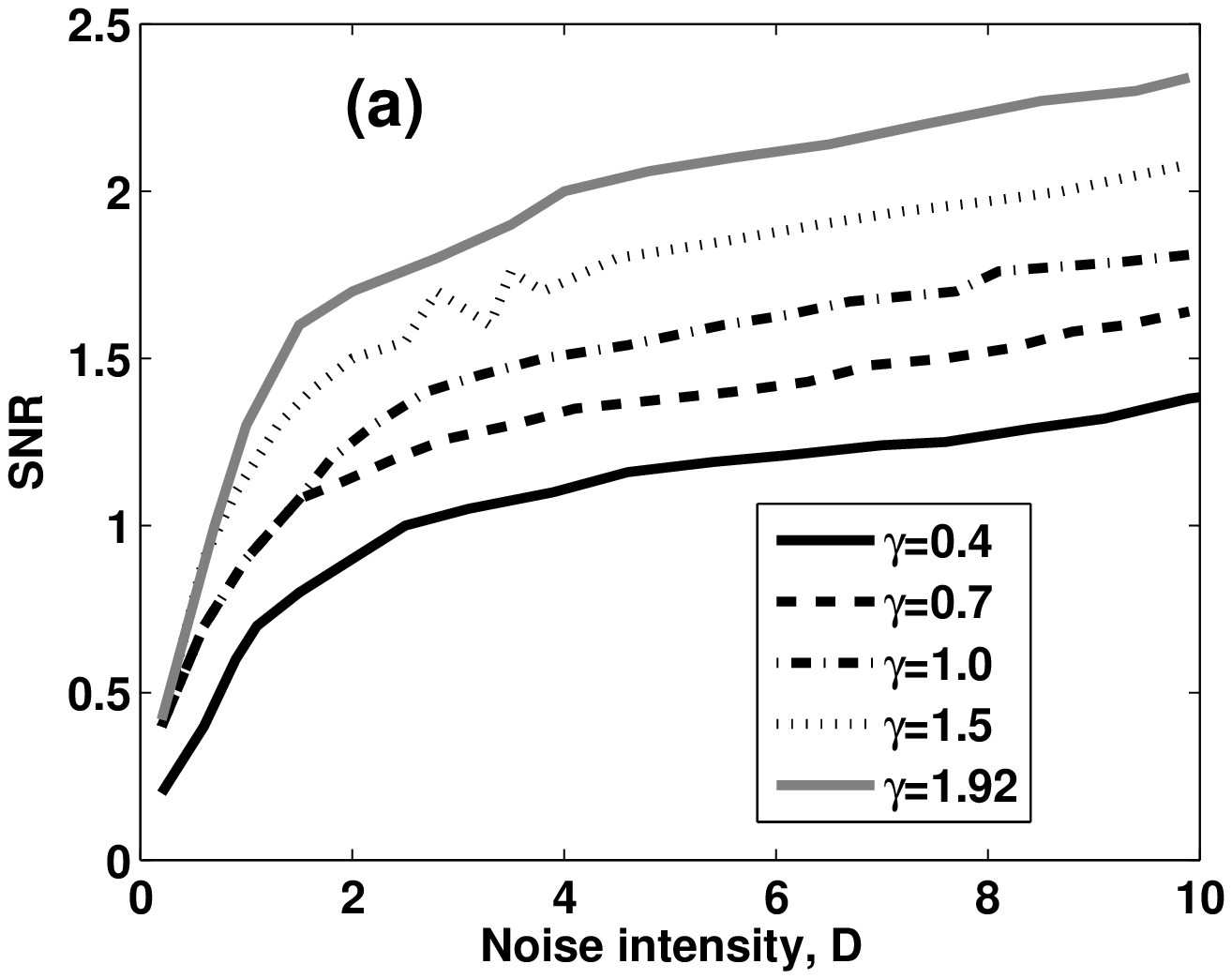}
\includegraphics[height=5.4cm,width=7.4cm]{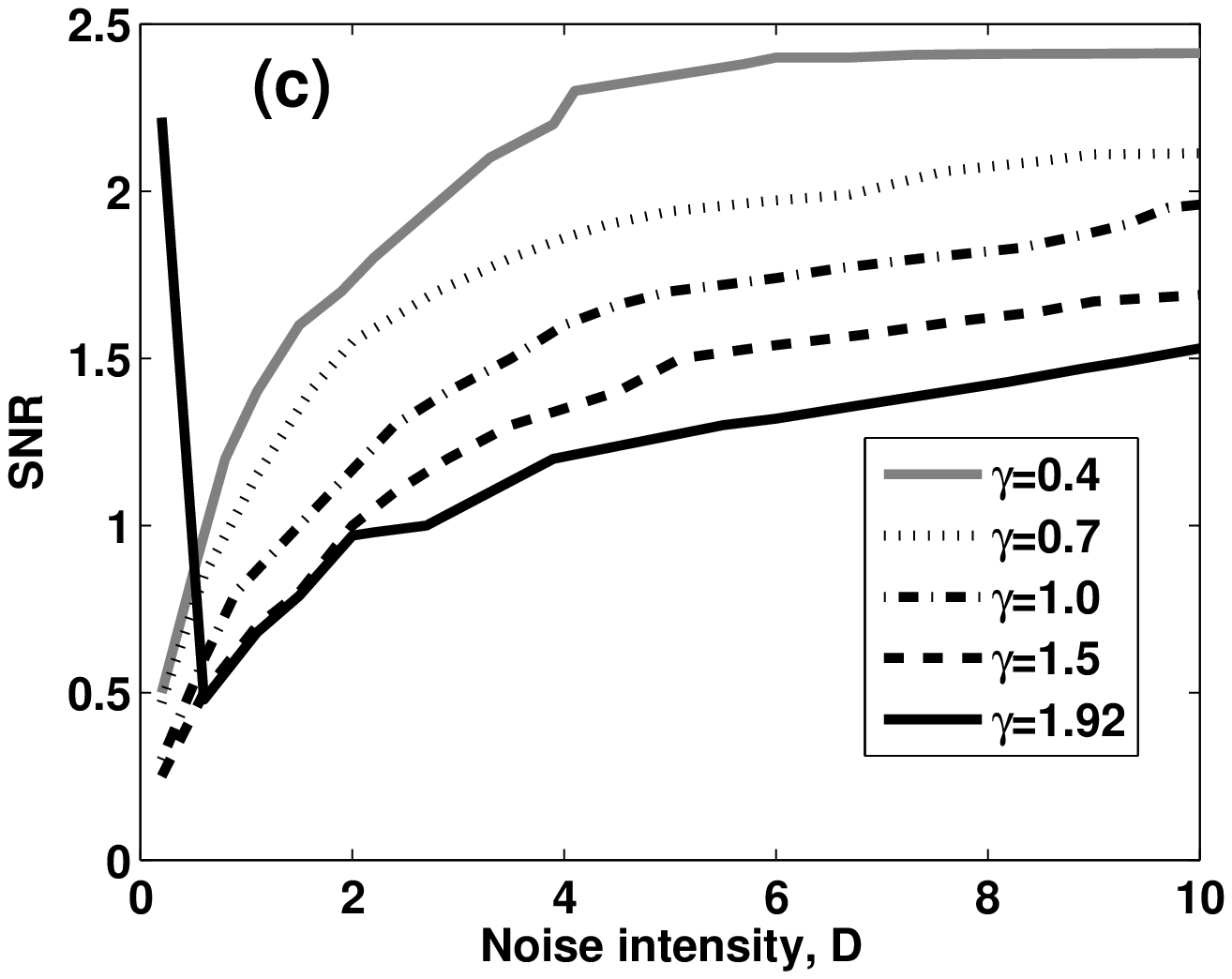}
\caption{\it
Effects of $\gamma$ on the variation of SNR versus $D$ for the asymmetric ($S_1$)
and symmetric ($S_3$) system, Table \ref{Tab1}. These figures correspond to
 the dynamical states when the initial conditions are choose on $A_3$ and
 the frequencies of the two stable attractors are both $\simeq 1.0$. All data refer to $\mu=0.01$, $E=0.25$, $\omega=0.5$. (a) correspond to ($0.114, 0.003$) and (b) correspond to ($0.12, 0.0016$).}
\label{fig7}
\end{center}
\end{figure}

The  SNR  for  different values of the control parameters $\alpha$ and $\beta$ and of the noise index  $\gamma$ is computed in Fig. \ref{fig7}, that  shows the SNR versus  $D$ for the asymmetric ($S_1$) and symmetric ($S_3$) system (the initial conditions are selected on the attractor $A_3$).
For few values of $\gamma$, from $0.4$ to $1.92$, the SNR increases with $D$, and no maximum is observed.
One concludes that presumably SR does not occur in this case.

\begin{figure}
\begin{center}
\includegraphics[height=5.4cm,width=7.4cm]{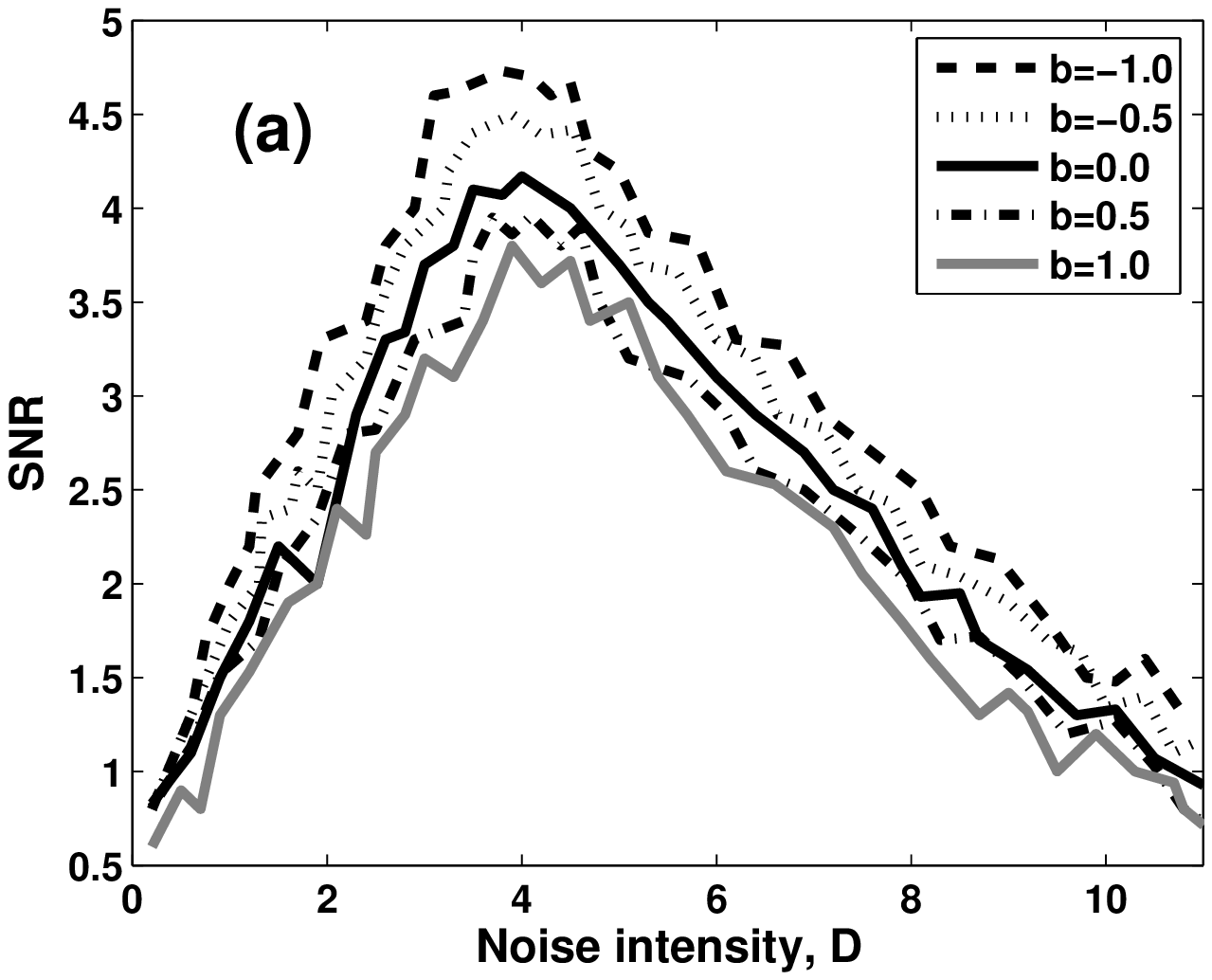}
\includegraphics[height=5.4cm,width=7.4cm]{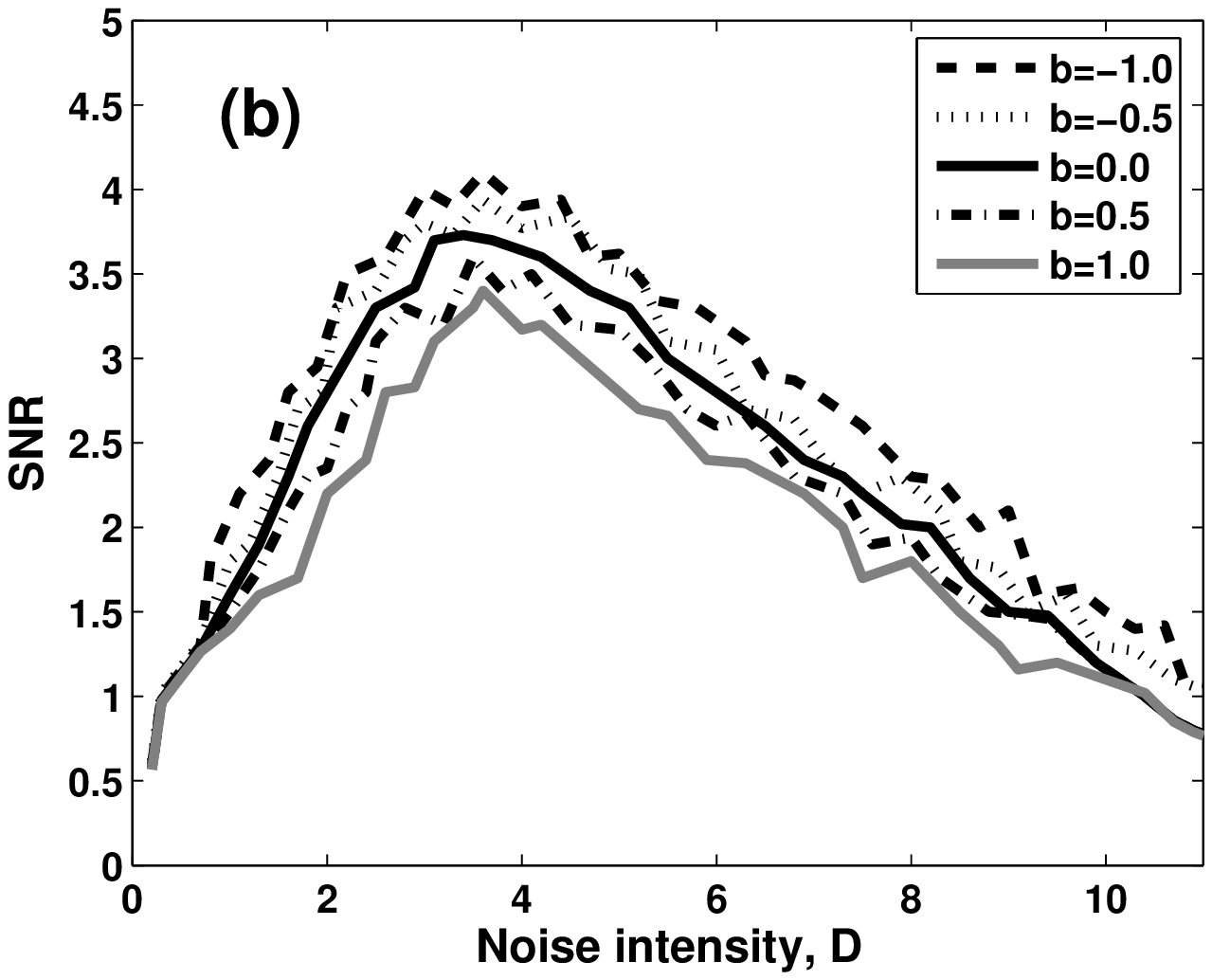}
\caption{\it
Effects of the skewness parameter $b$ and of the LN index $\gamma$ on the SNR versus $D$.
The panels correspond to the dynamical states resulting from initial conditions on the inner attractor $A_1$.
The frequencies of the two stable attractors are both $\simeq 1.0$.
Data refer to $(a)$: parameter $S_1$, $(b)$: parameters $S_3$ (see Table \ref{Tab1}).
The other parameters read: $E=0.25$, $\omega=0.5$, $\gamma=1.7.$ }
\label{fig8}
\end{center}
\end{figure}

\begin{figure}
\begin{center}
\includegraphics[height=5.4cm,width=7.4cm]{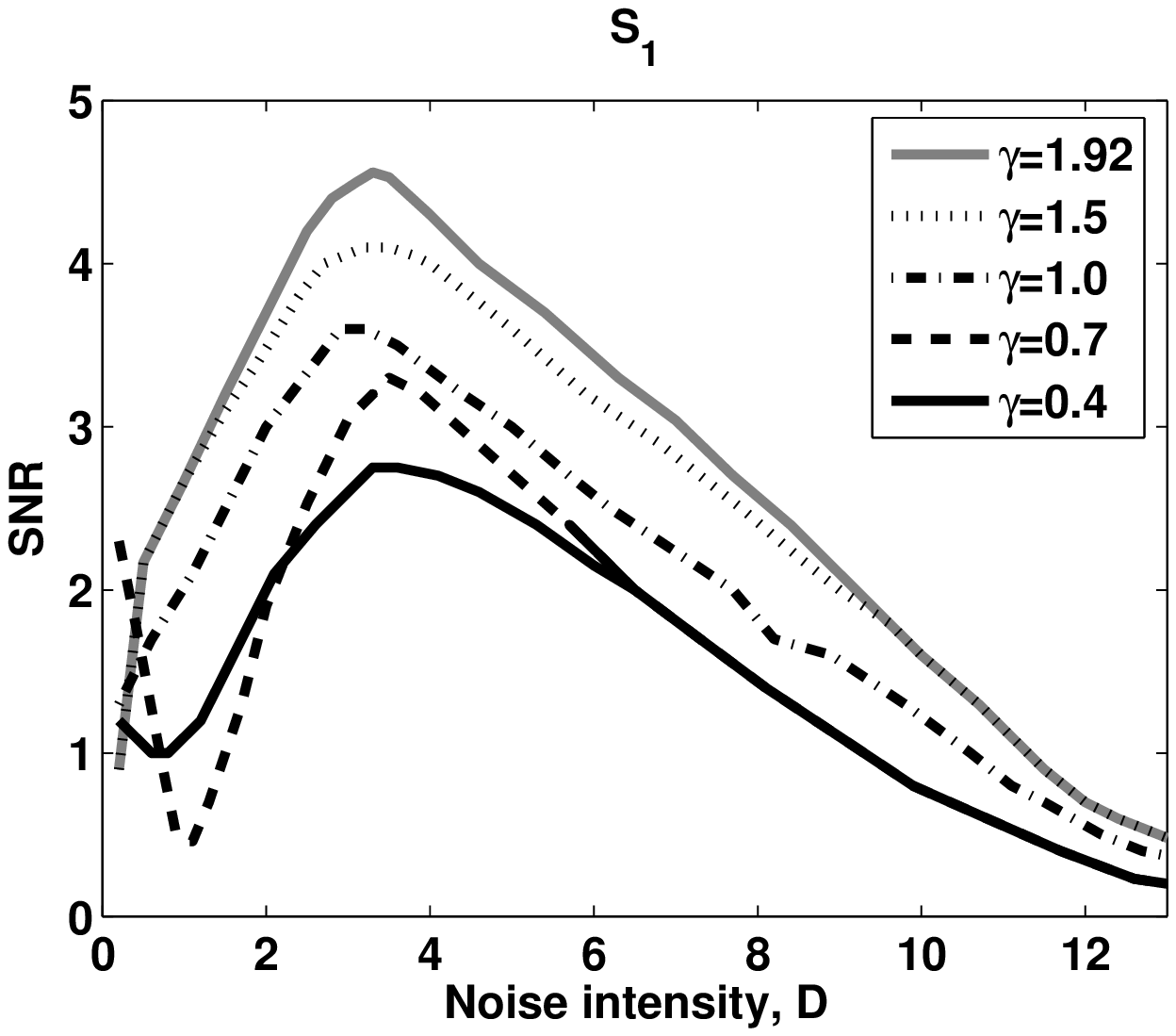}
\includegraphics[height=5.4cm,width=7.4cm]{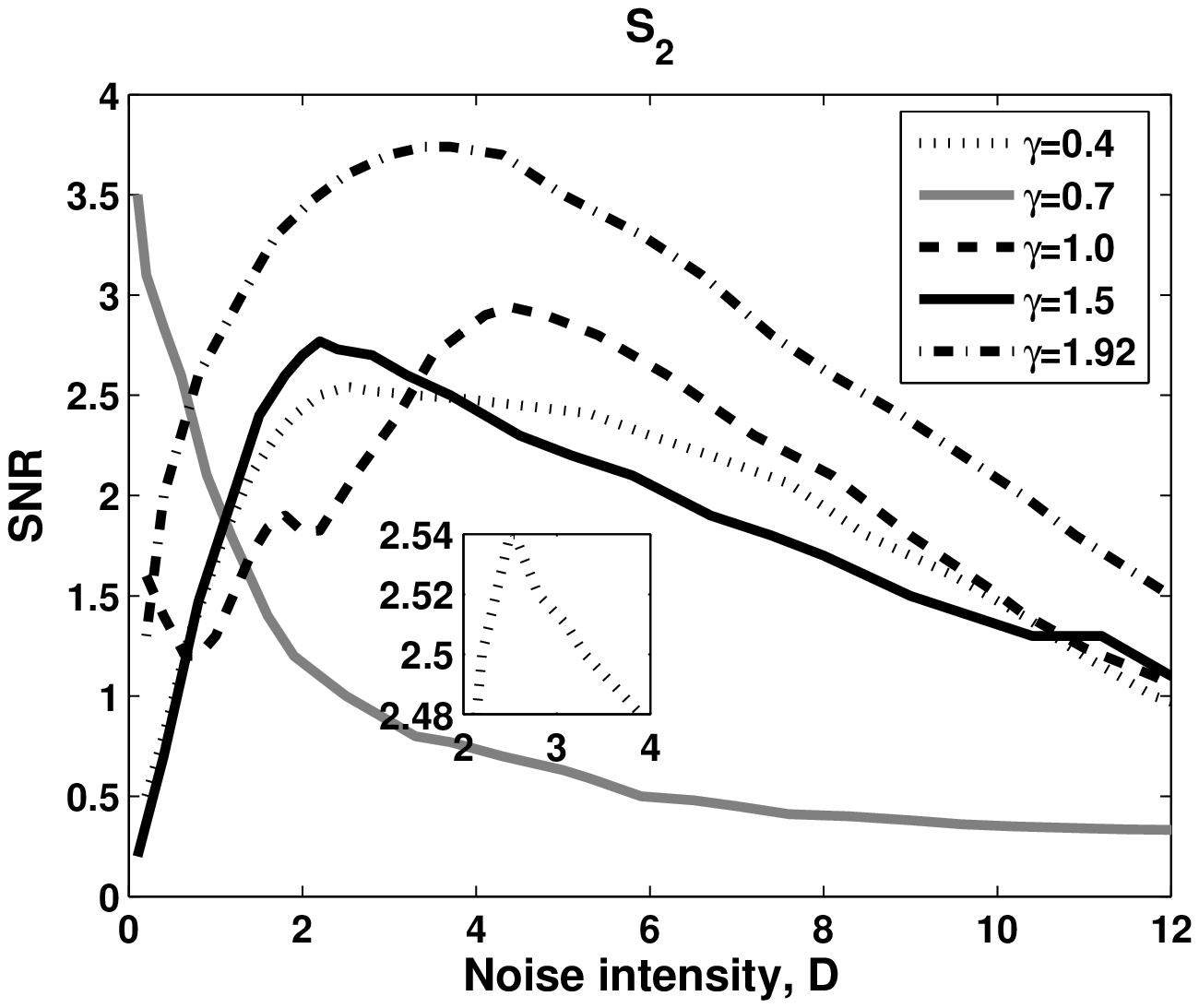}
\caption{\it
Effects of $\gamma$ on the variation of SNR versus $D$ for the asymmetric system.
These figures correspond to the dynamical states when the initial conditions
 are choose on $A_1$ and the frequencies of the two stable attractors are both
 $\simeq 1.0$. All data refer to $\mu=0.01$, $E=0.25$, $\omega=0.5$. }
\label{fig9}
\end{center}
\end{figure}

\begin{figure}
\begin{center}
\includegraphics[height=5.4cm,width=7.4cm]{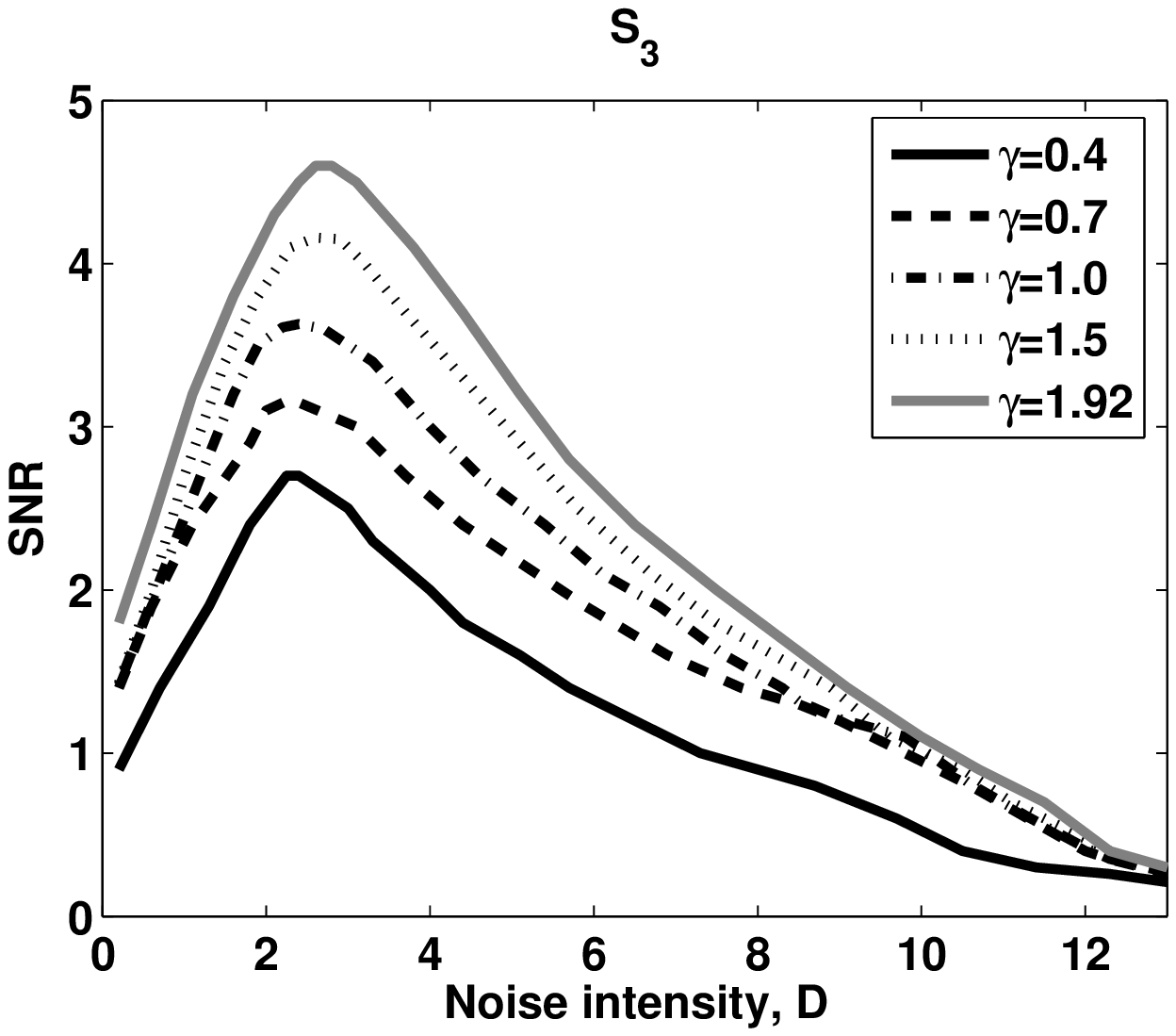}
\includegraphics[height=5.4cm,width=7.4cm]{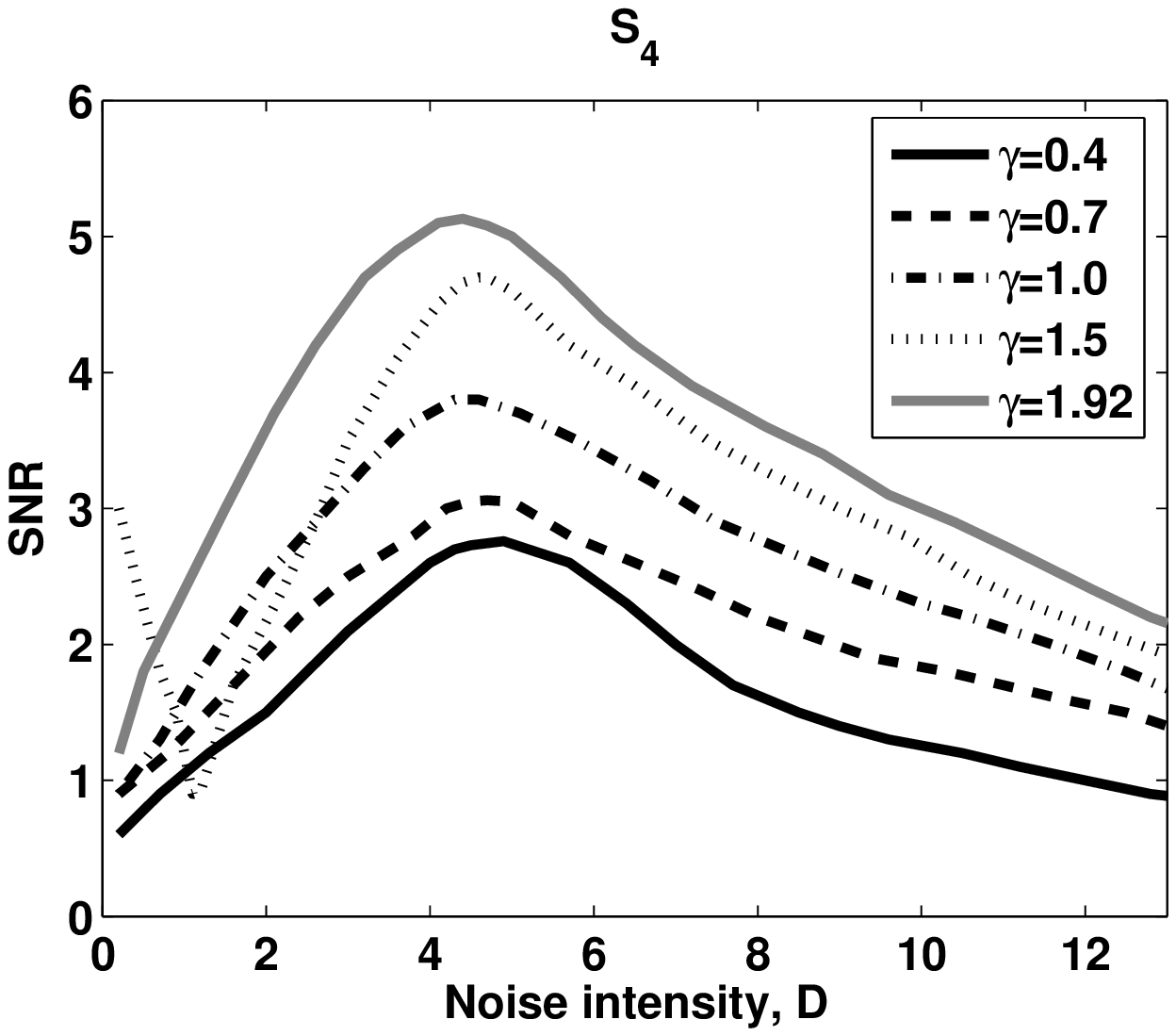}
\caption{\it
Effects of $\gamma$ on the variation of SNR versus  $D$ for the symmetric system.
These figures correspond to the dynamical states when the initial conditions
are choose on $A_1$ and the frequencies of the two stable attractors are both $\simeq 1.0$.
 All data refer to $\mu=0.01$, $E=0.25$, $\omega=0.5$. }
\label{fig10}
\end{center}
\end{figure}

\begin{figure}
\begin{center}
\includegraphics[height=5.4cm,width=7.4cm]{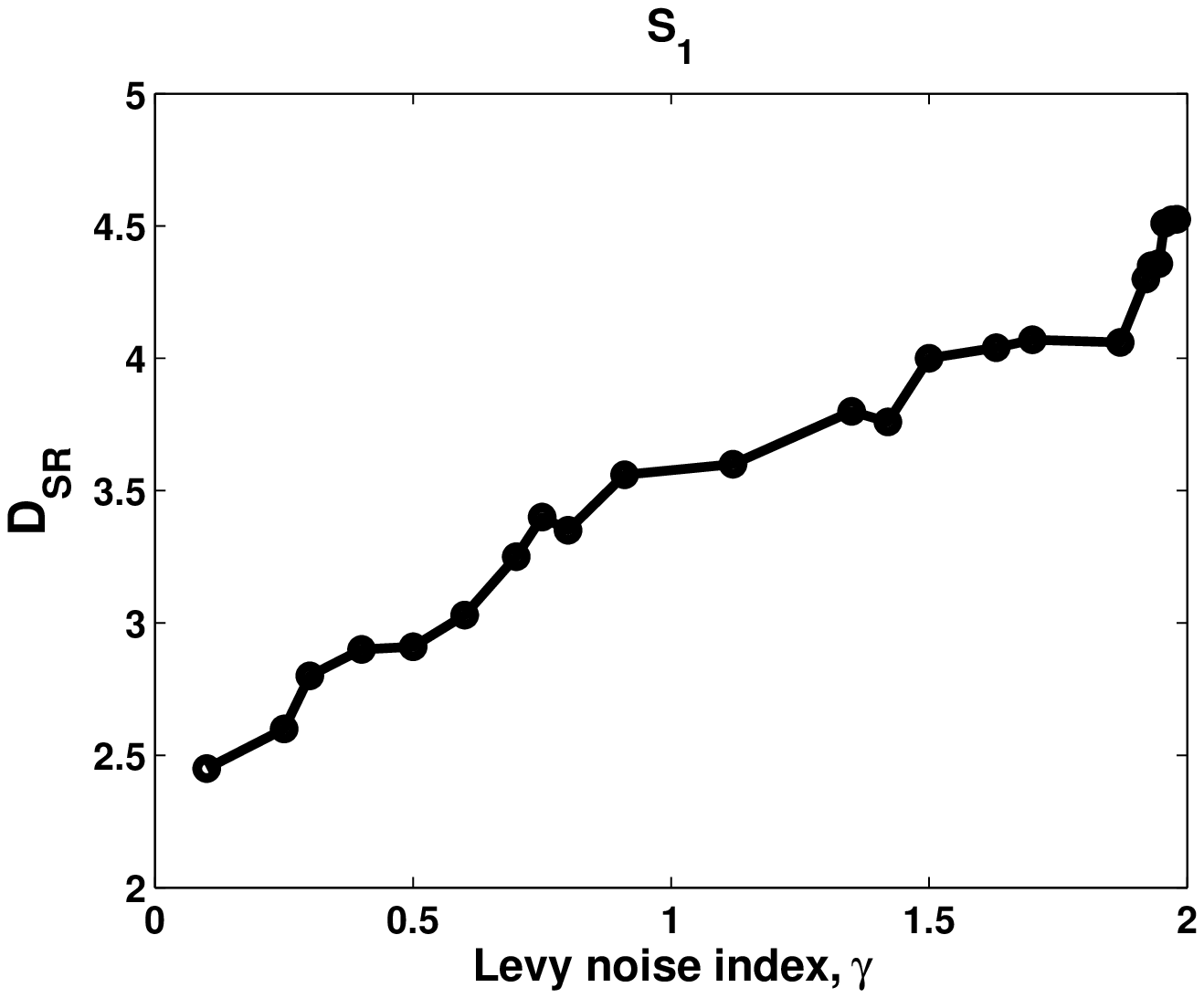}
\includegraphics[height=5.4cm,width=7.4cm]{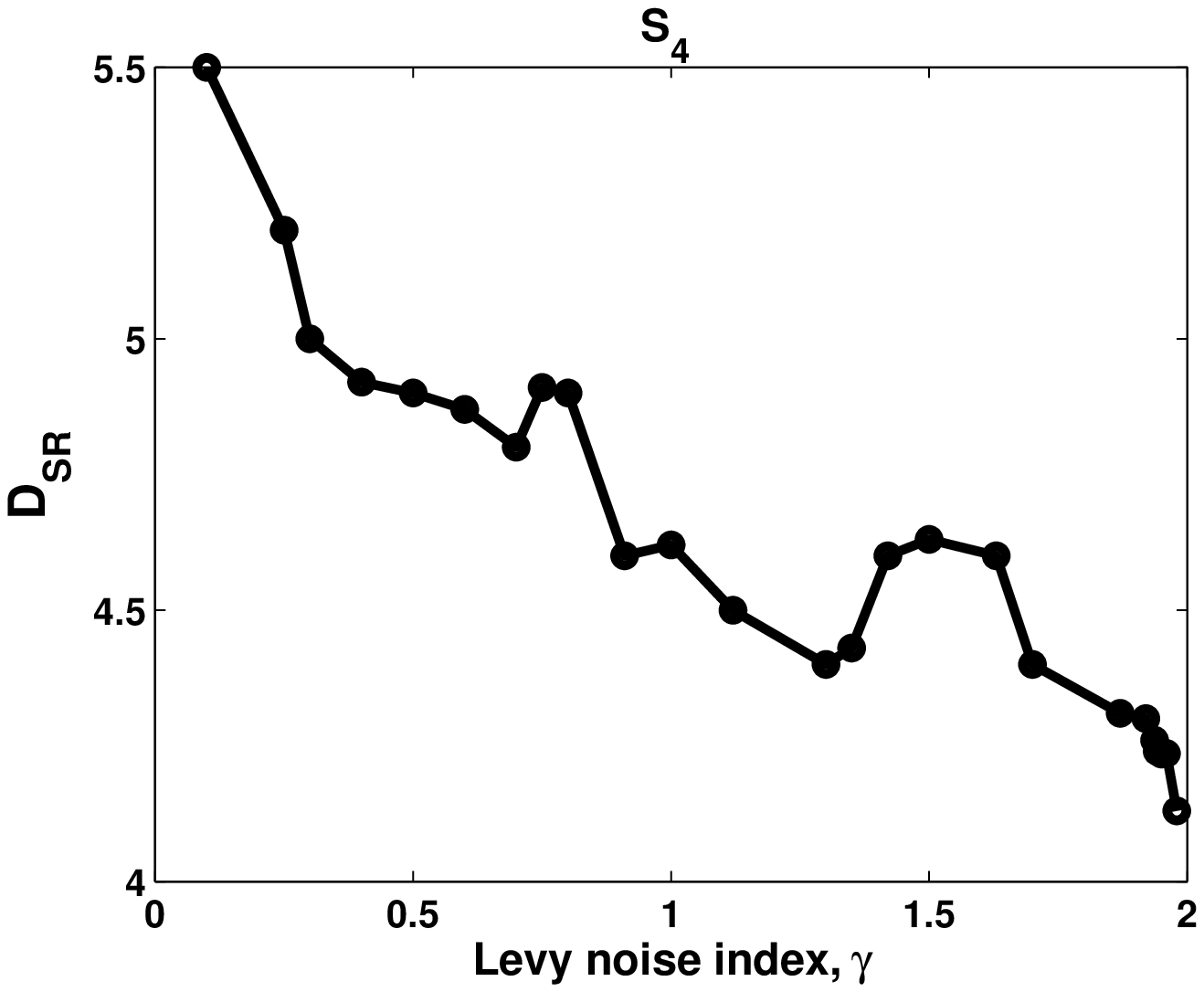}
\caption{\it
Variation of $D_{SR}$ (the  $D$ at which SNR is maximum)
versus $\gamma$ for initial conditions
choose on $A_1$ and for the set of parameters $S_1$ and $S_4$.
All data refer to $\mu=0.01$, $E=0.25$, $\omega=0.5$. }
\label{fig11}
\end{center}
\end{figure}

\begin{figure}
\begin{center}
\includegraphics[height=5.4cm,width=7.4cm]{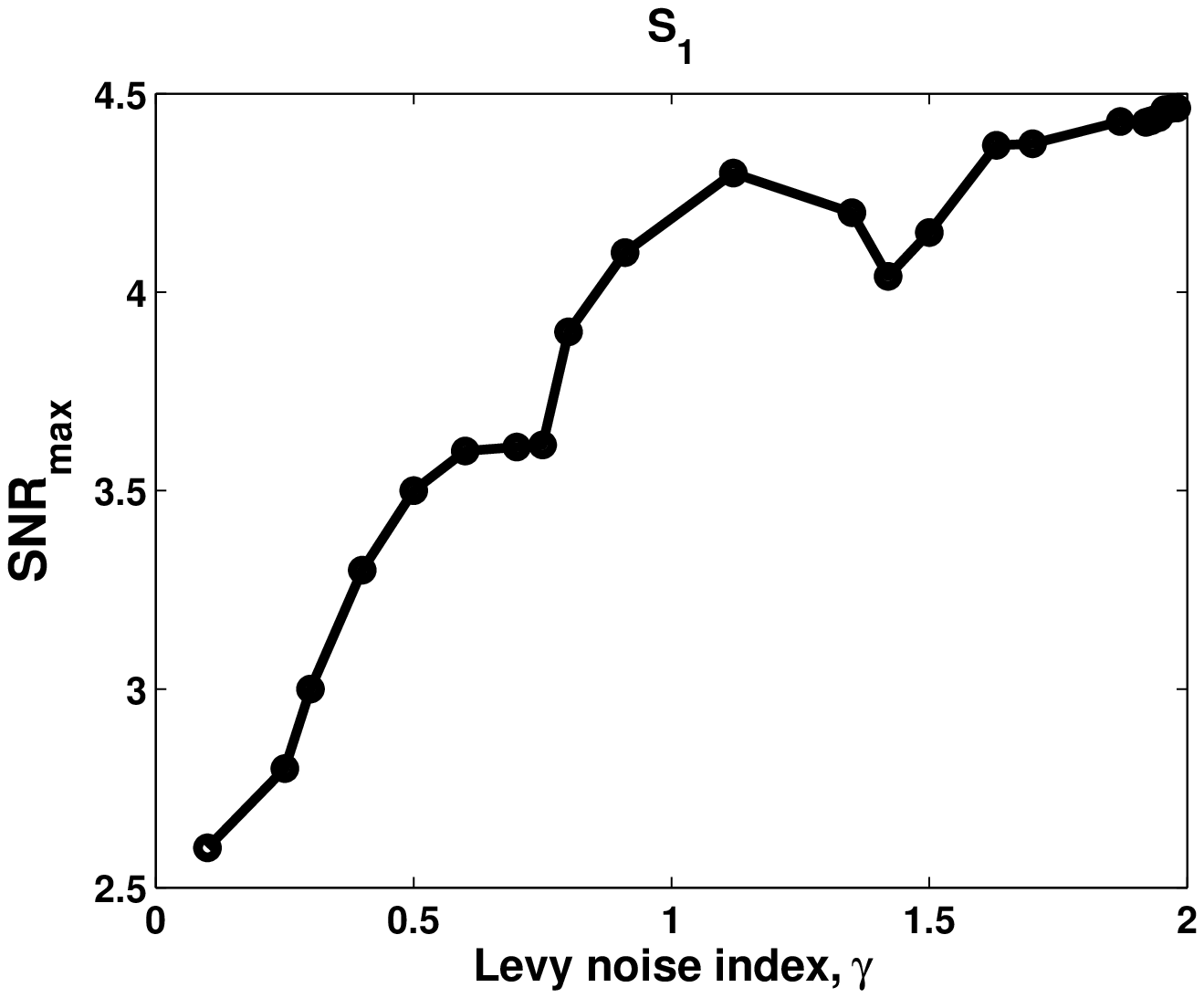}
\includegraphics[height=5.4cm,width=7.4cm]{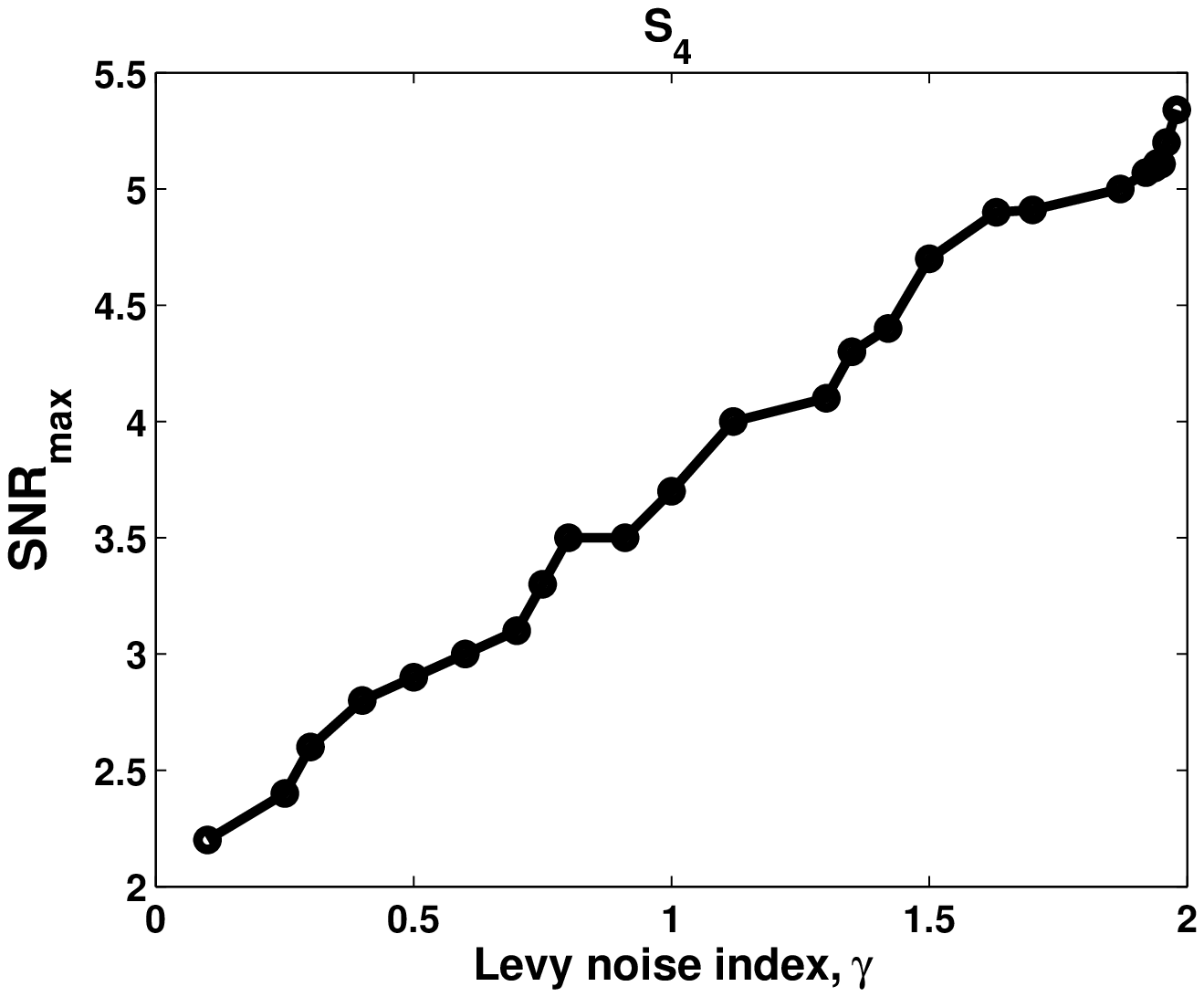}
\caption{\it
Variation of $SNR_{max}$ (the maximum of SNR) versus $\gamma$ for initial conditions choose on $A_1$ and for the
set of parameters $S_1$ and $S_4$ (see Table $1$ for details).
All data refer to $\mu=0.01$, $E=0.25$, $\omega=0.5$. }
\label{fig12}
\end{center}
\end{figure}

In  Figs. \ref{fig8} - \ref{fig10}, for initial conditions chosen on the inner $A_1$ attractor, it is shown the SNR versus $D$ for different values of the skewness parameter  $b$ (Fig.\ref{fig8}) and $\gamma$  (Figs. \ref{fig9} and \ref{fig10}).
One finds that the skewness parameter $b$ has a marginal effect, see Fig. \ref{fig8}.
It is evident that as the noise intensity $D$ increases, the SNR value also increases up to a maximum; such a noise intensity value  is named $D_{SR}$.
A further increase in the noise intensity $D$ leads to a decline of the SNR, that is precisely the signature of  the presence of SR.
In Figs. \ref{fig11} and \ref{fig12} are displayed the noise intensity $D_{SR}$ at which a peak of SNR the noise occurs and the amplitude of the SNR at the peak, respectively, as a function of the L\`evy noise index $\gamma$.
The  results of Fig. \ref{fig11}  indicate that increasing  the parameter $ \gamma $ the peak occurs at higher noise intensity for an attractor ($S_1$), and at a lower noise intensity for the other ($S_4$).
The   analysis  of  SNR for different $\gamma$ in Fig. \ref{fig12} indicates that the SNR curves have the same trend as a function of the parameter   $\gamma$
In general, thus, one can conclude that the peak value of SNR decreases and moves to larger noise intensity with the decrease of the index $\gamma$.
As a general conclusion about the role of the parameter $gamma$, we feel it appropriate to state that the data demonstrate a strong dependence of the SNR upon the L\`evy index $\gamma$.


\section{Conclusion}
\label{conclusion}

We have addressed the problem of the coherence and stochastic resonance in a non linear high order van der Pol system  driven by  L\`evy noise and a periodic signal.
Preliminarily, numerical simulations of stochastic trajectories under the effect of LN, demonstrated through several tools (ACF, PSD and SNR), indicate the existence of CR and SR of the system.
The influence of the  L\`evy index $\gamma$ on the feature of SR, measured by inspection of  SNR, in van der Pol birhythmic oscillator is influnced by the features of the double-well quasi- potential.
The behavior of the SR quantifiers is a clear indication that also  L\`evy noise, as ordinaty Gaussian noise, can co-operate with the spontaneous self sustaind cycles, as well as with the external periodic drive.
When SR occurs, the index $\gamma$ systematically leads to a decrease of the peak of the SNR; In these circumstances, it is possible to find a value of noise intensity, for which the SR phenomenon is observed.
In general, LN processes produce several continuous and spiking self sustained model describe by  modified van der Pol  because general forms of the SR "forbidden interval" theorem hold for several types of LN.

It is observed that the noise and the periodic drive affect the van der Pol's output power spectrum, that reaches a peak value at a certain frequency.
The results indicate that the decrease of  $\gamma$, that  leads to larger fluctuations and heavier tails of the noise, causes the switches between the two potentials wells to become more irregular, and consequently weakens the SR phenomenon.

However, for a certain fixed value of the L\`evy noise index $\gamma$, the birhythmic van der Pol systems' response is  analogous to ordinary bistable systems, and it is possible to retrieve the  influence  of  different  parameters  on  the optimal response (as measured by the SNR) as expected for the SR paradigm.

By way of conclusions, the mechanism of SR seems robust enough to be at work in spite of the contemporary presence of two variations respect to the standard case: non Gaussian and offers new perspectives on SR as a general phenomenon, that can be applied to such a diverse system.

\section*{Conflict of Interest}
The authors declare that they have no conflict of interest.

\section*{Author contribution statement}
This paper was proposed by  R. Yamapi and
G. Filatrella. R. Mbakob Yonkeu is the main author of
the paper, while J. Kurths was the supervisor of this
works.

\section*{Acknowledgments}
Part of this work was done during the visit of R.Yamapi at the Potsdam Institute for Climate Impact Research (PIK), Potsdam, Germany (September 28-October 18, 2019). RY acknowledges the support of the Institute.

\newpage

\newpage

\begin{thebibliography}{37}
\bibitem{pikovsky} Arkady S. Pikovsky and J Kurths, Coherence Resonance in a Noise-Driven Excitable System. Phys. Rev. Let. 78, 775-778 (1997).
\bibitem{Kenfack}A. Kenfack  and  K.Singh,  Stochastic  resonance  in  coupled  underdamped bistable systems. Phys. Rev. E {\bf 82}, 046224 (2010).
\bibitem{Alfonsi}L. Alfonsi,  L. Gammaitoni,  S. Santucci,  and  A. Bulsara,
``Intrawell  stochastic resonance versus interwell stochastic resonance in underdamped bistable systems''.
Phys. Rev. E {\bf 62}, 299 (2000).
\bibitem{Gu}R.Gu,  Y.Xu,  H.Zhang,  and  Z. Sun,
``Phase  transitions  and  the  mean  first passage  time  of  an  asymmetric  bistable  system  with  non-Gaussian  L\`evy  noise'',
 Acta Phys. Sin. {\bf 11}, 110514 (2011).
\bibitem{Zhao}Y. Zhao,  J. Li  and  X. Zhao,  Transition  and  transport  for  a  spatially  periodic stochastic  system  with  locally  coupled  oscillators.
Phys. Rev.  E {\bf 70},  031113 (2004).

\bibitem{Fiasconaro05} A. Fiasconaro, B. Spagnolo, and S. Boccaletti,
Signatures of noise-enhanced stability in metastable states,
Phys. Rev. E {\bf 72}, 061110 (2005).
\bibitem{Guarcello15} C. Guarcello, D. Valenti, and B. Spagnolo, Phase dynamics in graphene-based Josephson junctions in the presence of thermal and correlated fluctuations, Phys. Rev. B {\bf 92}, 174519 (2015).




\bibitem{Zakhrova} A. Zakharova,  T. Vadivasova,  V. Anishchenko,  and  J. Kurths,  Stochastic bifurcations  and  coherence-like  resonance  in  a  self-sustained  bistable  noise oscillator.
Phys. Rev. E {\bf 81}, 011106 (2010).
\bibitem{Mbakob2} R. Mbakob Yonkeu, R. Yamapi, G. Filatrella, and C. Tchawoua, Stochastic Bifurcations induced by correlated Noise in a Birhythmic van der Pol System, Comm. in Nonl. Sc. and Num. Sim. {\bf 33}, 70-84 (2016).
\bibitem{Mbakob3} R. Mbakob Yonkeu, R. Yamapi, G. Filatrella and C. Tchawoua, Effects of a Periodic Drive and Correlated Noise on Birhythmic van der Pol Systems, Physica A {\bf 466}, 552-569 (2017).
\bibitem{Chamgoue} R. Yamapi, A. Ch\'eag\'e Chamgou\'e, G. Filatrella, and P. Woafo, Coherence and stochastic resonance in a birhythmic van der Pol system.
Eur. Phys. J. B {\bf 90}, 153 (2017).

\bibitem{Biswas19} D. Biswas, T. Banerjee, and J. Kurths,
``Effect of filtered feedback on birhythmicity: Suppression of birhythmic oscillation'',
Phys. Rev. E {\bf 99}, 062210 (2019).


\bibitem{Benzi} R. Benzi, A. Sutera, and A. Vulpiani.
The mechanism of stochastic resonance.
J. Phys. A. {\bf 14}, L453-L457 (1981).

\bibitem{Mcdonnell}M. D. Mcdonnell, D. Abbott,
 What is stochastic resonance? Definitions, misconceptions, debates, and its relevance to biology.
Plos Computational Biology. {\bf 5(5)}, e1000348 (2009).

\bibitem{Gammaitoni} L. Gammaitoni, P. Hanggi, P. Jung, and F. Marchesoni.
Stochastic resonance.
Rev. Mod. Phys. {\bf 70}, 223-287 (1998).








\bibitem{H.Qiao}H.Qiao and J. Duan,
Asymptotic methods for stochastic dynamical systems with small non-Gaussian L\`evy noise.
Stochastics and Dynamics {\bf 15}, 1550004 (2015).

\bibitem{Bartlomiej}B. Dybiec and Ewa Gudowska-Nowak,
L\`evy stable noise-induced transitions:stochastic resonance, resonant activation and dynamic hysteresis,
J. Stat. Mech. P05004 (2009).

\bibitem{Guarcello13} C. Guarcello, D. Valenti, G. Augello, and B. Spagnolo,
The role of non-Gaussian sources in the transient dynamics of long Josephson junctions, Acta Phys. Pol. B {\bf 44}, 997 (2013).
\bibitem{Valenti14} D. Valenti, C. Guarcello, and B. Spagnolo,
Switching times in long-overlap Josephson junctions subject to thermal fluctuations and non-Gaussian noise sources,
Phys. Rev. B {\bf 89}, 214510 (2014).
\bibitem{Guarcello16}	C. Guarcello, D. Valenti, A. Carollo, and B. Spagnolo,
``Effects of L\`evy noise on the dynamics of sine-Gordon solitons in long Josephson junctions'',
J. Stat. Mech.: Theory Exp. 054012 (2016).



\bibitem{Guarcello17} C. Guarcello, D.  Valenti, b. Spagnolo, V. Pierro, and G. Filatrella,
``Anomalous transport effects on switching currents of graphene-based Josephson junctions''
Nanotechnology, {\bf 28}, 134001 (2017).
\bibitem{Guarcello19}C. Guarcello, D. Valenti, B. Spagnolo, V. Pierro, and G. Filatrella,
 ``Josephson-based Threshold Detector for Lévy-Distributed Current Fluctuations'',
Phys. Rev. Appl., {\bf 11}, 044078 (2019).



\bibitem{art}A. Patel and B. Kosko,
Stochastic Resonance in Continuous and Spiking Neuron Models with L\`evy Noise,
IEEE Trans. on Neur. Networks, {\bf 19}, 1993 (2008).

\bibitem{art1}N. Hohn and A. N. Burkitt,
Shot noise in leaky integrate-and-Fire neuron,
Phys. Rev. E {\bf 63},0319021 (2001).



\bibitem{B.D} B.D. Hughes, Random Walks and Random Environments (Oxford Science, Oxford, 1995), Vol. I
\bibitem{P.M}P.M. Drysdale and P.A. Robinson, L\`evy random walks in finite systems Phys. Rev. E {\bf 58}, 5382 (1998)
\bibitem{Anishchenko}V. S. Anishchenko,A. B. Neiman, F. Moss and L. Schimansky-Geier , 1992 Sov. Phys. Usp. 42 7
\bibitem{T.Srokowski}T.Srokowski,  Nonlinear  stochastic  equations  with
multiplicative  L\`evy  noise.Phys. Rev. E 81, 051110 (2010)

\bibitem{yamapi2019} R. Yamapi, R. Mbakob Yonkeu, G. Filatrella and J. Kurths,
L\`evy noise induced transitions and enhanced stability  in  a birhythmic van der Pol system.
Eur.  Phys.  J. B 92, 152(2019)

\bibitem{Dybiec}B. Dybiec and E. Gudowska-Nowak,
Stochastic resonance: the role of  alpha-stable  noises.
Acta Physica Polonica B {\bf 37}, 1479 (2006)

\bibitem{Reinker}S. Reinker, Stochastic resonance in thalamic neurons and resonant neuron models[D]. New York: The University of British Columbia, 2004


\bibitem{Frohlich1}H. Frohlich, Long range coherence and energy storage in a biological sys-tems. Int. J. Quantum Chem. {\bf 641}, 649-52 (1968).

\bibitem{Frohlich2}H. Frohlich In: Marois, editor, Quantum mechanical concepts in biology, in theoretical physics and biology, p. 13-22 (1969).

\bibitem{Yamapi10} R. Yamapi, G. Filatrella, and M. A. Aziz-Alaoui,
Global stability analysis of birhythmicity in a self-sustained oscillator,
Chaos {\bf 20}, 013114 (2010).

\bibitem{Ghosh11} P. Ghosh, S. Sen, S. S. Riaz, and D. S. Ray,
``Controlling birhythmicity in a self-sustained oscillator by time-delayed feedback'',
Phys. Rev. E {\bf 83}, 036205 (2011).

\bibitem{Guo19} Q. Guo, Z. Sun, W. Xu,
``Bifurcations in a fractional birhythmic biological system with time delay'',
Commun. Nonlinear Sci. Numer. Simulat. {\bf  72}, 318  (2019).


\bibitem{Enjieu}H.G. Enjieu Kadji, J.B. Chabi Orou, R. Yamapi, P. Woafo, Nonlinear dynamics and strange attractors in the
biological system, Chaos, Solitons Fractals, 32, 862--882 (2007).

\bibitem{Mbakob1}. R. Mbakob Yonkeu, R. Yamapi, G. Filatrella and C. Tchawoua, Quasipotential of birhythmic van der Pol type systems with correlated noise, Nonlinear Dynamics 84(2), 627 (2016).
\bibitem{Liang Y}Y. J. Liang, W.Chen: A survey on computing L\`evy stable distributions and a new MATLAB toolbox. Signal Processing. {\bf 93}, 242 (2013).
\bibitem{Applebaum}D. Applebaum, L\`evy Processes and Stochastic Calculus, Cambridge University Press, Cambridge (2004).
\bibitem{Dubkov}Dubkov A. A., Spagnolo B.: Langevin Approach to L\`evy flights in fixed potentials: Exact results for stationary probability distributions. Acta Physica Polonica B {\bf 38}, 1745 (2007).
\bibitem{Zheng} Y. Zheng, L. Serdukova, J. Duan and J. Kurths: Transitions in a genetic transcriptional regulatory system under L\`evy motion. Scientific Reports {\bf 6}, 29274 (2016).



\bibitem{Janicki}A. Janicki, Numerical and Statistical Approximation of Stochastic Differential Equations with Non Gaussian Measures, HSC Monograph, Wroc law, 1996

%

\bibitem{Stratonovich}R.L. Stratonovich. Topics in the theory of random noise vol. II. New York: Gordon and Breach; 1967.

\bibitem{Zhang and Song}L.  Zhang,  A.  Song,  and  J.  He,  Stochastic  resonance  of  a  subdiffusive  bistable  system  driven  by  L\`evy  noise  based  on  the  subordination  process.  J.  Phys.  A: Math. Theor. 44 , 089501 (2011).

\bibitem{Mbakob4}R. Yamapi, R. Mbakob Yonkeu, G. Filatrella and C. Tchawoua, Effects of noise correlation on the resonances of a van der Pol type birhythmic system, Comm. in Nonl. Sci. and Num. Sim. {\bf 33}, 70 (2018).












\end{thebibliography}
\end{document}